\documentclass{aa}  

\usepackage{graphicx}
\usepackage{xcolor}
\usepackage{txfonts}
\usepackage{float}
\usepackage[breaklinks, colorlinks, citecolor=blue, linkcolor=blue]{hyperref}

\renewcommand{\arraystretch}{1.2}

\begin{document}

   \title{Two domains of extended Lyman $\alpha$ emission around galaxies:\\from local radiation to environmental regulation}
   
   \titlerunning{Two domains of extended Ly$\alpha$ emission}
   \authorrunning{D. Kozlova et al.}

    \author{D. Kozlova \inst{1} \corrauth{dkozlova@aip.de}
            \and
          L. Wisotzki \inst{1} \email{lwisotzki@aip.de}
          \and
          J. Pharo \inst{1} \email{jpharo@aip.de}
          \and
          T. Urrutia \inst{1} \email{turrutia@aip.de}
          \and
          R. Augustin \inst{1} \email{raugustin@aip.de}
          \and
          Y. Guo \inst{2} \email{yuchengg@asu.edu}
          \and
          H. Kusakabe \inst{3} \email{haruka.kusakabe.takeishi@gmail.com}
          \and
          J. Schaye \inst{4} \email{schaye@strw.leidenuniv.nl}
          \and
          D. Smirnov \inst{1} \email{dsmirnov@aip.de}          
          \and
          I. Pessa \inst{1} \email{ipessa@aip.de}
          }

   \institute{%
   Leibniz-Institut für Astrophysik Potsdam (AIP), An der Sternwarte 16, 14482 Potsdam, Germany \\
         \and
    School of Earth \& Space Exploration, Arizona State University, 781 Terrace Mall, Tempe, AZ 85287, USA\\
    \and 
    Department of General Systems Studies, Graduate School of Arts and Sciences, The University of Tokyo, 3-8-1 Komaba, Meguro-ku, Tokyo, 153-8902, Japan \\
    \and 
    Leiden Observatory, Leiden University, PO Box 9513, 2300 RA Leiden, the Netherlands
     \\
             }

   \date{Received date / Accepted date }

  \abstract{We examine the relation between extended Ly$\alpha$ halos around high-redshift galaxies and the main factors responsible for driving the emission in such halos, in particular at distances around and beyond one virial radius $r_\mathrm{vir}$. To reach the required surface brightness sensitivity we take advantage of the MUSE eXtremely Deep Field (MXDF) survey, allowing us to probe levels as faint as $\sim 10^{-20}$ erg cm$^{-2}$ s$^{-1}$ arcsec$^{-2}$ in individual Ly$\alpha$ halos. Our sample consists of the 21 apparently core- and halo-brightest (yet intrinsically low luminosity $\log_{10}$L$_{\mathrm{Ly}\alpha} < 42.3$~erg~s$^{-1}$) Ly$\alpha$ emitters (LAEs) in the MXDF at $3<z<4$, with typical virial radii around 20~kpc. We measure their radial surface brightness profiles out to 50~kpc (more than $2r_{\mathrm{vir}}$) and investigate the correlations between surface brightness and internal (star formation rates of the host galaxies, SFR) or external influences (environmental density, $\delta+1$). We find a clear break in these correlations at radii around or just below $1r_{\mathrm{vir}}$. Below this break the emission correlates tightly with SFR (as expected) and not at all with $\delta+1$. Beyond $\sim 1r_\mathrm{vir}$(20~kpc) we observe the opposite trend with no dependence on SFR, but an emerging correlation with $\delta+1$. We compare our measurements with the expected integrated surface brightness from ultrafaint, individually undetected LAEs and find that the latter is insufficient to drive the observed correlation. We conclude that Ly$\alpha$ emission from the outer halos is regulated by the surrounding environment, but originates mostly from diffuse gas rather than discrete sources. 
}


  \maketitle
  \nolinenumbers

\section{Introduction}\label{section_intro}

Extended \ion{H}{I} Ly$\alpha$ emission is now detected around most star-forming galaxies at high redshifts \citep[e.g.][]{Hayashino_2004, Steidel_2011, Wisotzki_2016, Kusakabe_2022}, providing a valuable tracer of the cool gas in the circumgalactic medium (CGM). Yet the mechanisms responsible for this emission are still a matter of active debate. These complications arise from the diversity of physical processes in the CGM \citep[current state of art and challenges are greatly presented in reviews][]{Tumlinson_2017, Faucher_Giguere_2023}, from the resonant nature of the Ly$\alpha$ transition \citep[e.g][]{Osterbrock_1962, Adams_1972, Loeb_Rybicki_1999, Zheng_2002, Verhamme_2006, Dijkstra_Kramer_2012, Gronke_2017, Chung_2019, Li_Gronke_2022, Khoraminezhad_2025}, and also observationally from the extremely low surface brightness of circumgalactic Ly$\alpha$ emission \citep[e.g.][]{Hayes_2014, Wisotzki_2016, Leclercq_2017, Lujan_Niemeyer_2022, Erb_2023}.

By definition the CGM bridges scales from the interstellar to the intergalactic medium, implying that it harbours a broad range of ecosystems \citep{Tumlinson_2017}. The innermost CGM interfaces directly to the ISM and is in permanent exchange of material and energy with it. This transition zone can however be studied in detail only in the local Universe with sufficient resolution. At higher redshifts our picture of the CGM is more limited, and while the physical processes probably differ between the ``inner'' and ``outer'' CGM, there is no well-defined boundary between the two domains. Circumgalactic Ly$\alpha$ emission (often called Ly$\alpha$ halos) is mostly observed from the inner CGM of Ly$\alpha$ emitters (LAEs), with exponential surface brightness profiles of scale lengths of typically a few kpc -- several times the half-light radii of the host galaxies but only a small fraction of the expected virial radii \citep{Momose_2014, Leclercq_2017, Claeyssens_2022}. Ly$\alpha$ halos are observed around UV-bright Lyman Break Galaxies \citep[e.g.][]{Steidel_2011, Kusakabe_2022} and are more extended around Active Galactic Nuclei \citep[e.g.][]{Steidel_2000, Cantalupo_2014, Borisova_2016}, but these are usually much more massive systems so that the relative scaling with virial radius is not very different.

The established standard view is that most of the Ly$\alpha$ emission from the inner halos of non-AGN galaxies is powered by H-ionizing photons emitted by massive stars and subsequently modified by radiative transfer effects, in particular resonant scattering off neutral H atoms and absorption by dust grains. This scenario is supported by the fact that numerical simulations with Ly$\alpha$ radiative transfer are now able to reproduce the main observed properties of Ly$\alpha$ halos quite well \citep{Mitchell_2021, Byrohl_2021, Blaizot_2023}. 

The situation becomes much more uncertain in the outer CGM towards and beyond the virial radius. While these regions still emit measurable Ly$\alpha$ emission, it is extremely faint and has so far been traced mainly through stacking the profiles or images of hundreds to thousands of LAEs \citep{Matsuda_2012, Wisotzki_2018, Lujan_Niemeyer_2022, Kikuchihara_2022, Guo_2024_SB_profiles}. A common feature noted in several recent studies is an apparent flattening of the radial surface brightness profiles at radii $\ga 30$--50~kpc, corresponding to roughly 1--2~$ r_{\text{vir}}$ for the typically low stellar masses of such systems (e.g. \citealt{Guo_2024_SB_profiles}). One possible explanation for such a break is a transition from one halo term emission (arising from the CGM of the central galaxy) to a superposition of the contributions from surrounding external LAEs (i.e.\ the two halo term contribution to the observed surface brightness), which in a stack of many randomly oriented objects would be azimuthally averaged and appear indistinguishable from a large diffuse halo. This interpretation is supported by cosmological simulations with Ly$\alpha$ radiative transfer \citep{Zheng_2011, Lake_2015, Mitchell_2021, Byrohl_2021, Li_2025} as well as semianalytic models \citep{Mas_Ribas_2017, Bacon_2021} which typically find that the two halo term becomes highly relevant at about such distances. Quantitatively this effect is however still very uncertain as it depends strongly on the luminosity function of faint LAEs \citep[as discussed in detail by][]{Herrero_Alonso_2023} which is still poorly constrained by both observations and simulations.

Other possible contributions to Ly$\alpha$ emission at large distances from galaxies could be due to recombination radiation or scattered light from cool gas clouds at or outside the virial radius, powered by the local UV radiation field. While this emission is expected to be extremely faint if only the metagalactic UV background is considered, the local radiation field and thus also the Ly$\alpha$ surface brightness could be significantly boosted by nearby ``other'' star-forming galaxies or AGN \citep{Cantalupo_2005, Kollmeier_2010, Gallego_2021}. Maybe even the combined emission (including scattered radiation) from denser regions in the diffuse intergalactic medium could become detectable at sufficiently large radii \citep[e.g.][]{Bacon_2021, Byrohl_Nelson_2023}.

We adopt an observational approach to explore the properties of Ly$\alpha$ emission at large distances from host galaxies. Taking advantage of the exceptionally deep MUSE eXtremely Deep Field \citep[MXDF; ][]{Bacon_2023} data obtained with the MUSE instrument at the ESO-VLT, we study the phenomenology of ``outer halo'' emission on an object-by-object basis. Our data are deep enough to detect Ly$\alpha$ emission at radii comparable to the virial radii of our predominantly low-mass galaxies, allowing us to go beyond average sample properties (accessible via stacking) and investigate the actual diversity of phenomena encountered in this radial range. 

The paper is organized as follows: In Sect.~\ref{section_data} we characterize the data and the sample selection, followed in Sect.~\ref{section_analysis} by a description of the main data analysis steps, in particular the extraction and modelling of surface brightness profiles.  Section~\ref{section_results} documents our results: In Sect.~\ref{discrete_neighbors} we focus on the relevance of external (two halo term) objects for the observed Ly$\alpha$ emission, and in Sect.~\ref{UV_ionisation} we consider the impact of different contributions to the local UV radiation field. The implications of these results are discussed in Sect.~\ref{section_discussion}. We present our conclusions in Sect.~\ref{section_conclusions}. 

We assume a $\Lambda$CDM cosmology with $H_0 = 70$~km~s~$^{-1}$~Mpc~$^{-1}$, $\Omega_m = 0.3$, and $\Omega_{\Lambda} = 0.7$. All distances are expressed in physical units.

\section{Data and sample} 
\label{section_data}
    
\begin{figure}
    \centering
    \includegraphics[width=\hsize]{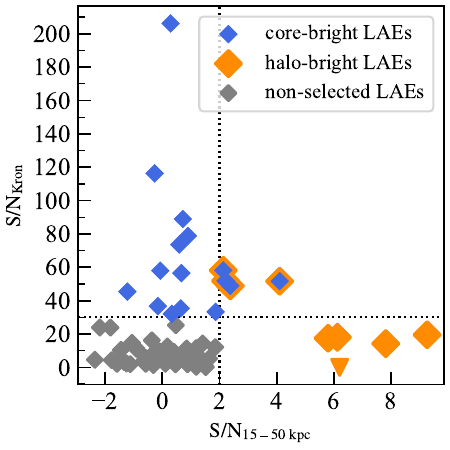}
    \caption{Ly$\alpha$ central (1 Kron radius) versus annular ($15<r<50$~kpc) signal-to-noise ratios of the parent sample (grey diamonds) and of the selected LAEs (orange filled and blue empty diamonds for halo- and core-bright LAEs, respectively). The dashed lines mark our selection thresholds of S/N$_{15-50 \text{kpc} \ge 2}$ and S/N$_{\text{Kron}} \ge 30$. Objects falling into both categories are marked accordingly. The orange triangle indicate ID~8537 with centrally suppressed Ly$\alpha$ emission for which no meaningful central S/N value could be measured.
    }
    \label{Fig_sample_in_MXDF}
\end{figure}

\subsection{Observational data}

Our main data source is the MUSE eXtremely Deep Field (MXDF) obtained with the Multi-Unit Spectroscopic Explorer (MUSE; \citealt{Bacon_2010}) instrument at the ESO Very Large Telescope. The observations were conducted in Wide-Field Mode enhanced by Ground-Layer Adaptive Optics (GLAO). MUSE is an integral-field spectrograph that delivers datacubes of a fixed spectral range from 4700~\AA\ to 9350\AA\ with a spectral resolution (FWHM) of $\sim$2.8\AA\ and a spatial GLAO-corrected point-spread function (PSF) of $\sim$0\farcs5 (FWHM) at 0\farcs2 spatial sampling. The MXDF consists of 141 hours of coadded MUSE exposures targeting a single pointing in the Hubble Ultra-Deep Field (HUDF). Because of multiple rotations between exposures the footprint of the combined MXDF dataset is nearly circular in shape, with the full-exposure region spanning a solid angle of 1\arcmin\ in diameter. A detailed description of the observations, data reduction, and the overall content of the survey can be found in \citealt{Bacon_2023}; the entire dataset with catalogs and derived quantities has been publicly released as Advanced MUSE Data Products (AMUSED)
\footnote{https://amused.univ-lyon1.fr/project/UDF/}.
We will refer to this dataset several times in the following and refer to individual objects by their AMUSED catalog identifiers.

As described below, we extract pseudo-narrowband images from the datacube that cover only the Ly$\alpha$ lines of certain sources. The achieved sensitivity depends on the wavelength (i.e.\ on the redshift of the object) and on the spectral width of the line; for an unresolved emission line the typical 1$\sigma$ surface brightness limit within an aperture of 1\arcsec\ is $10^{-19}$ erg s$^{-1}$ cm$^{-2}$ \citep{Bacon_2023}. Even lower surface brightness levels can be reached by averaging over larger apertures.

The MXDF is surrounded on all sides by the so-called MUSE-MOSAIC, a $3' \times 3'$ region with contiguous MUSE coverage at $10^\mathrm{h}$ exposure time \citep{Bacon_2017}. We employ the updated catalog of LAEs in the full MOSAIC (also provided via the AMUSED database) to characterize the environment of the objects in our subsample (see Sect.~\ref{discrete_neighbors_env}).

We also make use of the publicly available deep HST images in the HUDF/MXDF footprint (F775W band), obtained as object-centered cutout images from the AMUSED database.

\subsection{Sample selection and extraction}  
\label{sample}

In this work we study galaxies in the redshift range $2.9 < z < 4$ that show prominent Ly$\alpha$ emission, i.e. LAEs. Note that our LAE definition does not specify a minimum Ly$\alpha$ equivalent width, the only requirement is that the object is detected with MUSE by its Ly$\alpha$ line. Our parent dataset is drawn from the publicly released AMUSED catalog of the MXDF \citep{Bacon_2023}. Here we focus exclusively on a small subset from this parent sample with the highest signal-to-noise ratio in the Ly$\alpha$ line. Our galaxies are divided into two subsamples: (i) A ``core-bright'' sample of LAEs based on the Ly$\alpha$ emission from their central regions irrespective of the prominence of their extended halos, and (ii) an additional ``halo-bright'' sample specifically selecting objects with very extended Ly$\alpha$ emission detected at large distances. Below we explain in detail and quantify the selection criteria. 

We emphasize that although the galaxies in our two subsamples stand out as the brightest Ly$\alpha$ emitters in the MXDF, most of them have Ly$\alpha$ luminosities lower than $10^{42}$ erg s$^{-1}$, well below the sensitivity threshold of all but the deepest photometric narrowband imaging surveys and almost an order of magnitude fainter than $L^\star$ in the Schechter function parametrization of the Ly$\alpha$ luminosity function \citep{Herenz_2019}. These objects are ``bright'' only by their rank in the exceptionally deep MXDF, but not special otherwise; in conventional terms they even represent the category of ``faint'' LAEs.

The AMUSED catalog contains 664 galaxies. Restricting the search to objects with secure redshifts (confidence classes 2 and 3) reduces this to 503 entries. Because the exposure time in the MXDF decreases in the outer regions, we select only the 359 objects that received at least 100$^\mathrm{h}$ at their spatial centers. After limiting the selection to sources with Ly$\alpha$ detected in emission within our predefined redshift range, we are left with 217 galaxies.

For each of these objects we extract 21\arcsec $\times$ 21\arcsec\ pseudo-narrowband Ly$\alpha$ subimages from the continuum-subtracted datacubes (henceforth denoted as NB images), as described in detail in Sect.~\ref{pseudoNB_images} below. For objects for which these images extend outside the 100$^\mathrm{h}$ exposure time footprint we demand that at least 85\% of the NB image area must be contained within the footprint. This then reduces the MXDF parent sample to 81 entries.

\subsubsection{Core-bright Ly$\alpha$ emitters} 
\label{selection_bright_LAEs}

We define as core-bright all galaxies with central Ly$\alpha$ emission above a minimum flux, irrespective of their extended halo properties. This selection is representative of the larger parent LAE population. We employ the LSDCat code \citep{Herenz_2017}, specifically the \texttt{lsdcat\_measure} routine, to perform adaptive aperture photometry in the NB images, where the apertures are determined from the data as multiples of 3-dimensional Kron radii (see \citealt{Herenz_2017} for the precise definition of these radii). We adopt a radius of 1~$r_{\text{Kron}}$, which captures the LAE central region but leaves out most of the extended halo. As selection criterion we adopt a signal-to-noise ratio (S/N) in this aperture of $\text{S/N}_{\text{Kron}} > 30$.

By visual inspection we find that in some objects this flux might be affected by contamination due to other sources. To avoid such contamination, we carefully mask the region around each object to remove bright continuum objects (which sometimes leave continuum subtraction residuals in the NB images) as well as other sources of line emission (Ly$\alpha$ or other) appearing at these wavelengths. In some cases of neighboring same-redshift galaxies that might be surrounded by their own extended low-level Ly$\alpha$ halo emission we fit and subtract a 2-dimensional surface brightness model of the neighbor to remove also its possible extended halo (details of the modeling procedure are given in Sect.~\ref{galfit_modeling}). A few other cases are individually discarded from the sample because of compromised data, such as one source where the redshifted Ly$\alpha$ happens to be affected by full-frame $[$\ion{O}{iii}$]\lambda5007\AA$ emission at $z=0$ originating in the ISM of the Milky Way. Our final core-bright LAE sample consists of 16 objects.

\subsubsection{Halo-bright Ly$\alpha$ emitters} 
\label{selection_bright_LAHs}

Since in this study we are specifically interested in the outermost regions of Ly$\alpha$ halos, we supplement the core-bright sample with objects selected by the condition that they show significantly detected extended emission at very large radii. This sample is therefore falls outside of the representation of general LAEs population but adds valuable information to the ``outer halo'' characterization. We define an ``outer halo'' region by two circular contours at fixed physical separation. The inner radius of $r_{\mathrm{i}}=15$~kpc roughly corresponds to 2\arcsec\ at these redshifts and includes essentially all of the emission from the central galaxy (also taking PSF blurring into account) as well as most of the inner circumgalactic zone. The outer circle with $r_{\mathrm{o}}=50$~kpc extends out to, and in most cases even beyond, the typical virial radii ($\simeq$20--30~kpc) of low-luminosity LAEs at these redshifts (e.g. \citealt{Leclercq_2017, Herrero_Alonso_2023, Guo_2024_spectra}). The Ly$\alpha$ flux from within the annulus between $r_{\mathrm{i}}$ and $r_{\mathrm{o}}$ therefore originates mostly from the outer CGM.

We then select all objects with detected Ly$\alpha$ emission in this 15--50~kpc annulus. We integrate in the extracted NB images the total signal within the annulus, as well as its statistical uncertainty. For the latter we use the empirically rescaled ``effective noise'' \citep{Urrutia_2019} that accounts for covariances in the resampled MUSE datacube \citep{Weilbacher_2020} as well as potential noise enhancements due to small-scale systematics; see \citealt{Bacon_2023}). Only objects with $\text{S/N}_{15-50 \text{kpc}} > 2$ are admitted into the final ``halo-bright'' sample. The resulting sample of halo-bright LAEs consists of 9 objects of which 4 are already selected as ``core-bright''. 

Figure~\ref{Fig_sample_in_MXDF} shows how the two different S/N measurements relate to each other.
Notably, the fluxes within the central Kron radius and the outer halo aperture show essentially no correlation. In this figure we introduce a color code to distinguish between the two selection approaches that we maintain throughout this paper, adopting blue for the core-bright and orange for the halo-bright objects, respectively. The basic properties of our LAE samples are summarized in Table~\ref{table:1}.

Some of the halo-bright LAEs need a separate discussion as their morphological structures deviate from the typical LAH (Ly$\alpha$ halo) shape of a bright compact center surrounded by an extended halo that gets dimmer outwards. ID~8537 has a relatively bright UV continuum counterpart in the HST image, but exhibits a central depression in its Ly$\alpha$ emission at the location of the continuum source. Its LAH is very extended and approximately ring-shaped, although of clumpy appearance (see \citealt{Kusakabe_2022} for a detailed analysis of the LAHs of this and similar objects). Another irregularity is ID~7586 with its neighboring LAE ID~8469, forming a same-redshift binary system that most likely shares a single Ly$\alpha$ halo. These objects require some small modifications in their analysis, which we discuss on a case-by-case basis in Appendix \ref{ap:complex_laes}. Another pair of objects, ID~103/ID~8332 in AMUSED, was also initially selected, but then discarded due to its complex environment with several LAEs at the same redshift, which makes disentangling individual SB profiles and spectra extremely challenging, in particular defining an ``outer halo'' region. This system was also discussed by \citet{Kusakabe_2022}.

\begin{table*}
\caption[]{Sample properties.}    
\label{table:1}      
\centering         
\begin{tabular}{l l l l l l l l l l} 
\hline\hline       
ID & RA & DEC & ${z_{\text{sys}}}$ & ${z_{\text{sys}}}$ ref & ${\text{log}_{10}\text{L}_{\text{Ly}{\alpha}}}$ & M${_{F775W}}$ & S/N${_{\text{Kron}}}$ & S/N${_{15-50\text{kpc}}}$ & Sample\\ 
\hline   
6700${^{\diamond}}$ & 53.16827 & $-$27.78104 & 2.995  & \ion{C}{iii} $\lambda$1909 & 42.28 & 25.67 & 206.2 & 0.3 & C\\
106${^{\diamond}}$ & 53.16376 & $-$27.77902 & 3.276 & \ion{C}{iii} $\lambda$1909 & 42.39 & 26.51 &  116.3 &  0.3 & C\\
180 & 53.16396 & $-$27.77967 & 3.455 & Ly$\alpha$ & 42.15 & 27.28 & 88.9 & 0.7 & C\\
6690 & 53.16009 & $-$27.78610 & 3.746 & Ly$\alpha$ & 42.04 & 28.67 &  78.8 &  0.9 & C\\
6298${^{\diamond}}$ & 53.16931 & $-$27.78125 & 3.134 & \ion{O}{iii} $\lambda$1663 & 41.94 & 27.98 &  73.7 &  0.6 & C\\
469 & 53.17082 & $-$27.78479 & 3.762 & Ly$\alpha$ & 41.78 & 29.04 &  57.9 &  ${-0.1}$ & C\\
6929 & 53.16224 & $-$27.77930 & 3.716 & Ly$\alpha$ & 41.83 & 27.24 &  56.5 &  0.7 & C\\
7187 & 53.16917 & $-$27.78925 & 3.037 & Ly$\alpha$ & 41.70 & 27.76 &  45.4 &  ${-1.2}$ & C\\
590 & 53.17013 & $-$27.78105 & 3.671 & Ly$\alpha$ & 41.50 & 29.51 &  36.8 &  ${-0.2}$ & C\\
305 & 53.16910 & $-$27.78099 & 3.041 & Ly$\alpha$ & 41.42 & 28.26 &  35.4 &  0.6 & C\\
8327 & 53.16519 & $-$27.78971 & 3.706 & Ly$\alpha$ & 41.56 & 28.54 &  33.3 &  1.9 & C\\
660 & 53.15797 & $-$27.78709 & 3.189 & Ly$\alpha$ & 41.52 & 29.85 &  32.1 &  0.3 & C\\

2726 & 53.17147 & $-$27.78487 & 3.134 & Ly$\alpha$ & 41.68 & 28.12 &  51.7 &  4.1 & H, C\\ 
4842 & 53.15911 & $-$27.78843 & 3.169 & Ly$\alpha$ & 41.63 & 29.59 &  48.9 &  2.4 & H, C\\    
400 & 53.16316 & $-$27.77910 &  3.086 & Ly$\alpha$ & 41.74 & 28.79 &  51.9 &  2.2 & H, C\\  
6297 & 53.15927 & $-$27.78497 & 3.693 & Ly$\alpha$ & 41.97 & 28.92 &  58.2 &  2.1 & H, C\\ 

7586 & 53.16721 & $-$27.78536 & 3.064 & Ly$\alpha$ & 41.39 & 29.72 & 19.6 &  9.3 & H\\
8360 & 53.16687 & $-$27.79005 & 3.066 & Ly$\alpha$ & 41.50 & 29.90 & 14.3 &  7.8 & H\\
8537${^{\diamond}}$ & 53.16965 & $-$27.78808 &  3.188 & \ion{Si}{ii} $\lambda$1527 &  39.92 & 26.07 &  &  6.2  & H\\
8377 & 53.16478 & $-$27.78793 & 2.994 & Ly$\alpha$ & 41.09 & 26.21 & 18.0 &  6.1 & H\\
8284 & 53.16814 & $-$27.79079 & 3.704 & Ly$\alpha$ & 41.25 & 30.78 &  17.5 &  5.8 & H\\ 
\end{tabular}
\tablefoot{(1) ID: running source ID number. (2), (3) RA, DEC: centroid of extracted Ly$\alpha$ pseudo-narrowband image (see Sect.~\ref{pseudoNB_images}) (4) ${z_{\text{sys}}}$ defined based on (5) reference line, for Ly$\alpha$ line ${z_{\text{sys}}}$ is defined using empirical recipes provided by \citealt{Verhamme_2018}. (6) $\log_{10}$L$_{\mathrm{Ly}\alpha}$ $[$erg s$^{-1}]$ Ly$\alpha$ luminosity is calculated using fluxes integrated over segmentation maps (see details in \citealt{Bacon_2023}). (7) M$_{F775W}$ is a magnitude of HST counterpart in F775W filter. Data for (1)$-$(5) is provided in \citealt{Bacon_2023}. (8) S/N$_{\text{Kron}}$: signal-to-noise ratio within 1~Kron radius. (9) S/N$_{15-10\mathrm{kpc}}$: signal-to-noise in area $15<r<50$~kpc around Ly$\alpha$ NB centroid. (10) Sample: classification of each object based on core(C)-/halo(H)-brightness.
}
\end{table*}

\section{Analysis} 
\label{section_analysis}

\subsection{Pseudo-narrowband Ly$\alpha$ images}
\label{pseudoNB_images}

\begin{figure*}
    \centering
    \includegraphics[width=\hsize]{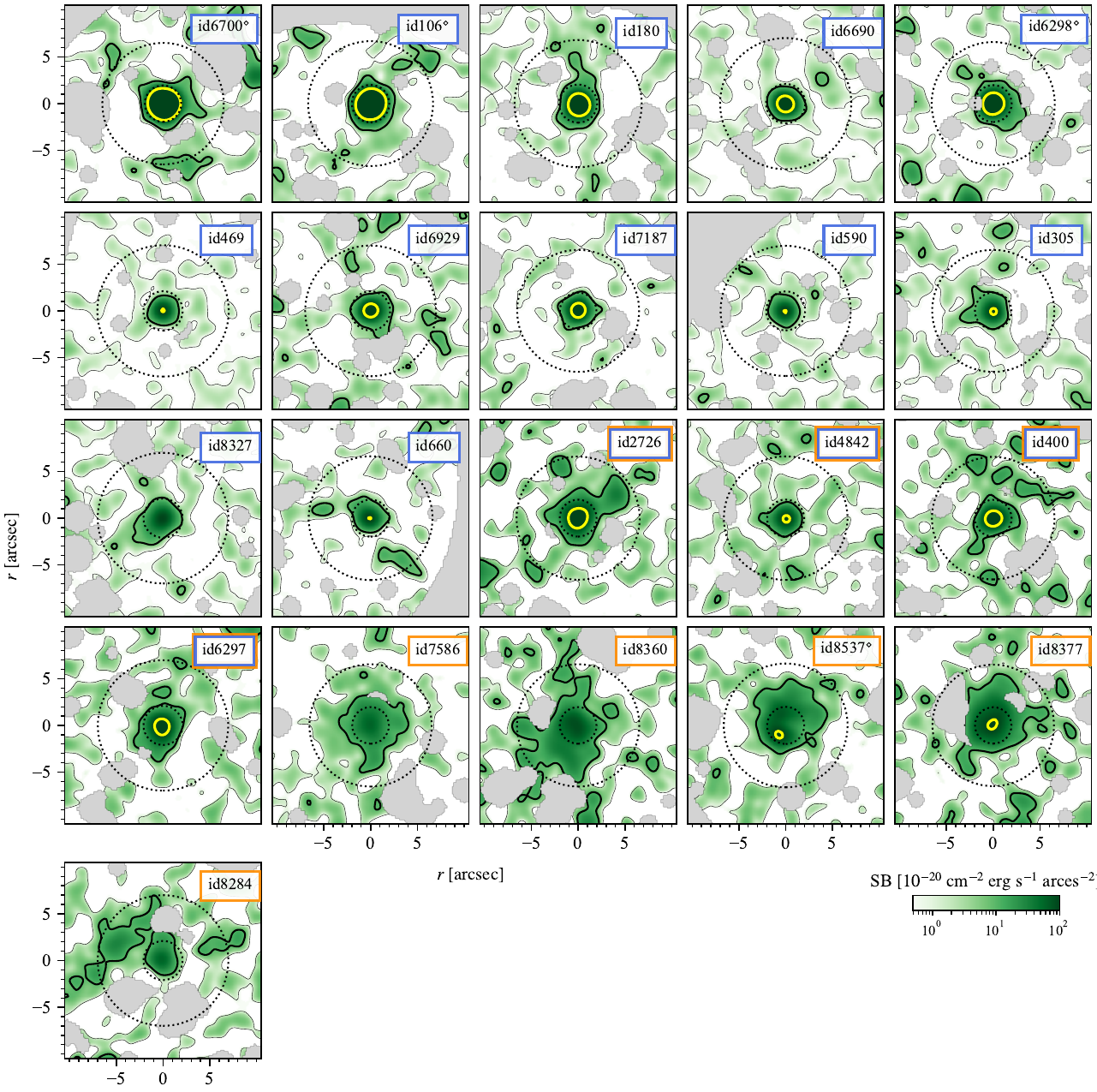}
    \caption{%
    Ly$\alpha$ pseudo-narrowband surface brightness images of all LAEs in the sample, smoothed by a Gaussian kernel with $\sigma = 0.75\arcsec$. The color of the box around each ID number indicates the subset to which the object belongs (orange: halo-bright, blue: core-bright). The objects are sorted in descending order by their Ly$\alpha$ signal-to-noise ratios, using $\mathrm{S/N}_{15-50 \mathrm{kpc}}$ for the halo-bright and $\mathrm{S/N}_{\mathrm{Kron}}$ for the core-bright LAEs. The black dotted circles mark radii of 15~kpc and 50~kpc, respectively. The contours (thin black, bold black, yellow) correspond to surface brightness levels of (1, 10, 100)~$\times$~$10^{-20}$~erg~cm$^{-2}$~s$^{-1}$~arcsec$^{-2}$. The gray patches indicate areas that were masked for the profile analysis and model fits.
    }
    \label{Fig_NB_images}
\end{figure*}

In order to extract the pseudo-narrowband (NB) Ly$\alpha$ images from the MUSE datacube, we set the corresponding spectral windows individually, taking the different emission line profiles into account. We determine the extraction bands for each object as follows:

\begin{itemize}
    \item In a first step we extract a provisional Ly$\alpha$ spectrum using the line peak and the spatial centroid from the parent AMUSED catalog. The spectrum is obtained by summing the flux for from the spaxels of the continuum-subtracted datacube over a square area of $5\arcsec\times5\arcsec$, for a fixed spectral range of $\pm 25$~\AA. 
    \item In each extracted spectrum we model the Ly$\alpha$ line by fitting a ``skewed Gaussian'' profile \citep{Shibuya_2014} to the data and parametrized in the same way as in \citet{Bacon_2023}:

    \begin{gather} \label{formula_skewed_gaus}
        f(\lambda; A, \mu, \sigma, \gamma) = \frac{A}{\sigma\sqrt{2\pi}} e^{[-(\lambda-\mu)^2/2\sigma^2]}\left[ 1 + \mathrm{erf} \left[  \frac{\gamma(\lambda-\mu)}{\sigma\sqrt{2}}\right ] \right]
    \end{gather}

    where $A$ is the amplitude, $\sigma$ the standard deviation, $\mu$ the center, $\gamma$ the so-called skewness parameter, and erf is the standard error function.  
    \item LAEs with a secondary peak (``blue bump'', either marked as such in AMUSED or judged visually by us) are modeled in a similar fashion, but using the combination of two skewed Gaussian distributions instead of one.
    \item We calculate the spectral curve of growth (COG) of the model spectrum.
    \item We define the blue and red boundaries of the adopted pseudo-narrowband as the wavelengths at which the COG reaches 2\% and 98\% of the total model line flux, respectively. Resulting pseudo-narrowband widths are $350<\Delta v < 2000$ km/s and allow for possible Ly$\alpha$ blueshift at $r>30$~kpc (described in \citealt{Guo_2024_spectra}. 
\end{itemize}

All NB images are then extracted by summing the datacube over the extent of the estimated individual spectral windows. An image with pixel-by-pixel uncertainties is created at the same time by standard error propagation from the ``effective variance'' cube. The resulting NB images of the LAEs in our sample are shown in Fig.~\ref{Fig_NB_images}, smoothed by a Gaussian kernel with $\sigma = 0.75\arcsec$ for display purposes.

\subsection{Two-dimensional modeling of morphological structures}
\label{galfit_modeling}

We approximate the Ly$\alpha$ surface brightness distribution by the sum of two spatial components, as introduced by \citet{Wisotzki_2016} and adopted by many authors since. The first (central) component is assumed to follow the UV continuum spatial shape and is described as an elongated exponential disk (or as a point source if the continuum counterpart is unresolved). The second (extended or ``halo'') component is also parameterized as an exponential (often circular) disk. We perform the modeling with the \texttt{galfit} code \citep{Peng_2002}, for which we take the NB image with its corresponding error image and the appropriate point spread function (PSF) as inputs. 

\subsubsection{Central component}
\label{galfit_modeling_central}

Following \citet{Wisotzki_2016}, we first perform a preparatory \texttt{galfit} modeling step of the UV continuum counterpart (as listed in the AMUSED database) to each Ly$\alpha$ source. We choose the HST F775W band which covers the spectral range longward of Ly$\alpha$ for our LAEs and provides the deepest and highest resolution counterpart information. The corresponding PSF is taken from \citet{Rafelski_2015}. We restrict the fits to a fitting area of $\sim$1\arcsec--1.5\arcsec, ensuring that only the LAE counterpart is modeled. The HST LAEs counterparts are mostly well-defined, with the only exceptions described in Appendix~\ref{ap:complex_laes}. 

We assume the UV continuum and the Ly$\alpha$ to be cospatial and thus fix the spatial centroid to that determined from the NB image, by fitting a 2D Gaussian using the \texttt{astropy.photutils} Python package \citep{Bradley_2024}. An exception is the centrally absorbed object ID~8537, for which we adopt the centroid of the UV counterpart for the compact Ly$\alpha$ component. Another special case is ID~106, whose HST counterpart has a complex morphological structure suggestive of a merging system of two galaxies connected by a tidal feature. For this object we model the HST counterpart as the superposition of two separate objects. We nevertheless still consider ID~106 as a single LAE, consistent with the finding by \citet{Vitte_2024} that both the blue and red peaks of its Ly$\alpha$ spectral profile emerge equally from the whole system.

All structural parameters obtained in the \texttt{galfit} model of the continuum counterpart (position angle PA, the minor-to-major axis ratio $q$, and the scale length $r_\mathrm{s,c}$) are then adopted as hard priors to model the central component in the Ly$\alpha$ NB images from MUSE. However, we also explore relaxing these priors by allowing $r_\mathrm{s,c}$ as a free parameter, to gain insight into how much these continuum priors influence the final Ly$\alpha$ model parameters. This test is documented in Appendix~\ref{ap:scalelengths}. We find that allowing $r_\mathrm{s,c}$ to float often gives physically unrealistic values of the central component scale length, but also that fixing or not fixing $r_\mathrm{s,c}$ has no significant influence on the parameter values of the extended halo component, especially its scale length $r_\mathrm{s, h}$. We therefore adopt the fit parameters based on fixed HST priors, leaving only the Ly$\alpha$ flux of the central component as a free parameter.

\subsubsection{Extended halo component}
\label{galfit_modeling_extended}

To fit the extended component we assume that the Ly$\alpha$ halo is approximately circular, restricting the model to only two free parameters (exponential scale length and the halo flux). We also explore models with elongated halo components, but find that these make essentially no difference for the azimuthally averaged surface brightness profiles which are the main focus of this study. This conclusion agress with studies of observed LAHs asymmetry, which suggest that they do not differ significantly from circular symmetry \citep{Chen_2021}.  On the other hand the shape parameters PA and $q$ come out as often poorly constrained, suggesting that the departure from circular symmetry cannot be large in those objects. The only exception is the likely merger ID~106 for which the halo is significantly elongated. 

Our final model for the Ly$\alpha$ emission is thus a sum of a central elongated and an extended circular exponential. There are six resulting fit parameters: From the fit to the HST counterpart image we obtain $r_\mathrm{s,c}$, $q_\mathrm{c}$, and PA$_{c}$. Modeling the Ly$\alpha$ NB image provides $r_\mathrm{s,h}$ and the halo flux to total flux ratio ${F_\mathrm{h}/F_\mathrm{tot}}$. The best-fit values of these parameters for our sample are listed in Table~\ref{table:2}. Errorbars are provided by \texttt{galfit}. 

A special treatment is required for ID~8537 with its centrally suppressed Ly$\alpha$ emission. Instead of the two-component model we mask out the central 2\arcsec and model only the extended component, taking the centroid from the UV counterpart as stated above, which prevents us from calculating central component parameters and ${F_\mathrm{h}/F_\mathrm{tot}}$ for this object.

\subsection{Radial surface brightness profiles}
\label{radial_profiles}

We extract radial profiles from the Ly$\alpha$ NB images by taking the azimuthal average of the surface brightness within concentric annuli. Out to a radius of 15~kpc we use a fixed width for each annulus of 1 MUSE spatial pixel (0\farcs2) to trace the inner halo regions most strongly affected by PSF blurring. For the outer halo regions at $r>15$~kpc we adopt the following larger widths of the annuli to increase the signal-to-noise ratio: $15.0<r<20.27$~kpc, $20.27<r<27.39$~kpc, $27.39<r<37.0$~kpc, and $37.0<r<50.0$~kpc. Uncertainty estimates for the radial profiles are calculated by standard error propagation from the NB image pixel variances. 

\begin{figure*}
    \centering
    \includegraphics[width=\hsize]{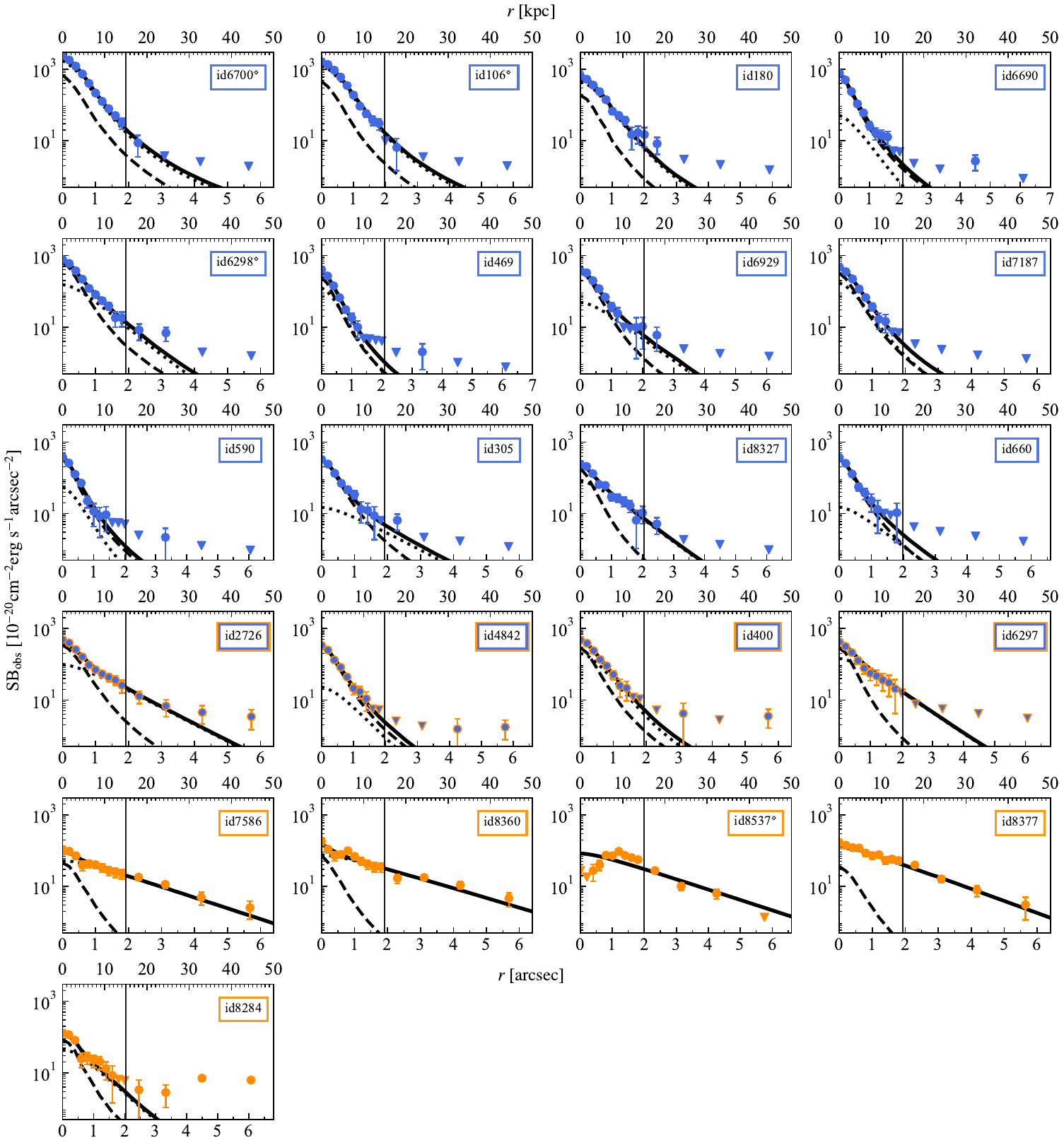}
    \caption{Surface brightness profiles of the LAEs in our sample (blue dots: core-bright,  orange: halo-bright LAEs). The solid lines show two-component \texttt{galfit} models, with the central (dashed line) and extended components (dotted). The vertical lines mark the physical radius of $r = 15$~kpc at the redshift of each object. SB$_\mathrm{obs}$ $1\sigma$ upper limits are indicated by triangles.
    }
    \label{Fig_SB_profiles}
\end{figure*}

The observed radial profiles, together with the smooth profiles obtained from the two-dimensional model fits, are presented in Fig.~\ref{Fig_SB_profiles}, with the same color coding as adopted before.

\begin{table*}
\caption{Model properties.}             
\label{table:2}      
\centering          
\begin{tabular}{l l l l l l l l l} 
\hline\hline       
ID & $r_\mathrm{s, c}$ & $q_\mathrm{c}$ & PA$_\mathrm{c}$ & $r_\mathrm{s, h}$ & $F_\mathrm{h}/F_{\mathrm{tot}}$ \\ 
\hline
6700 & 0.40$\pm$0.02 & 0.93$\pm$0.06 & 59.68$\pm$28.83 & 1.26$\pm$0.13 & 0.78 \\ 
106 & 0.39$\pm$0.10 & 0.53$\pm$0.23 & $-$20.00$\pm$24.27 & 2.08$\pm$0.06 & 0.38 \\
106 & 0.33$\pm$0.15 & 0.92$\pm$0.68 & 75.00$\pm$246.68 &  &  \\
180 & 0.45$\pm$0.09 & 0.60$\pm$0.16 & 20.00$\pm$22.72 & 1.81$\pm$0.20 & 0.81 \\
6690 & $-$ & $-$ & $-$ & 2.02$\pm$1.15 & 0.15 \\  
6298 & 0.28$\pm$0.15 & 0.77$\pm$0.63 & 83.86$\pm$101.56 & 3.63$\pm$0.52 & 0.49 \\  
469 & 0.16$\pm$0.08 & 1! & 0! & 1.08$\pm$0.61 & 0.42 \\  
6929 & 0.45$\pm$0.11 & 1! & 0! & 4.34$\pm$1.43 & 0.39 \\  
7187 & 0.47$\pm$0.77 & 0.50! & 38.75$\pm$114.26 & 1.48$\pm$0.78 & 0.47 \\  
590 & $-$ & $-$ & $-$ & 1.28$\pm$1.45 & 0.22 \\  
305 & $-$ & $-$ & $-$ & 6.51$\pm$3.98 & 0.27 \\  
8327 & 0.56$\pm$0.57 & 0.20! & $-$61.85$\pm$28.28 & 4.29$\pm$0.64 & 0.70 \\  
660 & $-$ & $-$ & $-$ & 4.13$\pm$4.18 & 0.18 \\ 
2726 & 0.70$\pm$0.14 & 0.36$\pm$0.17 & $-$45.00! & 6.44$\pm$0.79 & 0.63 \\  
4842 & 0.47$\pm$0.26 & 0.45$\pm$0.41 & $-$16.06$\pm$40.94 & 2.93$\pm$2.60 & 0.27 \\   
400 & 0.50$\pm$0.23 & 0.15$\pm$0.30 & 74.69$\pm$10.64 & 2.10$\pm$0.57 & 0.60 \\   
6297 & 0.45$\pm$0.09 & 0.51$\pm$0.19 & $-$29.76$\pm$16.67 & 4.86$\pm$0.40 & 0.76 \\ 
7586 & $-$ & $-$ & $-$ & 10.72$\pm$0.91 & 0.94 \\
8360 & $-$ & $-$ & $-$ & 12.05$\pm$0.93 & 0.96 \\
8537 &  &  &  & 10.83$\pm$0.60 &  \\
8377 & 0.97$\pm$0.07 & 0.31$\pm$0.05 & $\pm$24.85$\pm$6.20 & 9.58$\pm$0.55 & 0.97 \\  
8284 & $-$ & $-$ & $-$ & 3.41$\pm$1.58 & 0.64 \\  
     
\\ 
\end{tabular}
\tablefoot{ID: AMUSED identifier. $r_\mathrm{s, c}$: central component exponential scale length of HST UV continuum in proper kpc. $q_c$: minor to major axis ratio of the central component. PA$_\mathrm{c}$: position angle of the central component. $r_\mathrm{s, h}$: extended (halo) component exponential scale length in proper kpc. $F_\mathrm{h}/F_{\mathrm{tot}}$: integrated model halo Ly$\alpha$ flux to total integrated Ly$\alpha$ flux ratio calculated from model output magnitudes. Objects with an unresolved UV counterpart were modelled assuming a point source, so the central component parameters for these remain undefined and marked as $-$. Blank spaces correspond to absence of fit: see detailed discussion for ID~8537 and ID~106 in Sect.~\ref{galfit_modeling}. If a quantity was fixed for the fit at some value this is marked with ``!''. All halos are close to circular, except for ID~106 (which central component consists of two centrals surrounded by one halo) for which the additional parameters are $q_\mathrm{h} = 0.61\pm0.03$ and PA$_\mathrm{h} = -42\degr\pm 3\degr$.}
\end{table*}

\section{Results}
\label{section_results}

\subsection{Outer halo Ly$\alpha$ emission from individual objects}

\begin{figure}
    \centering
    \includegraphics[width=\hsize]{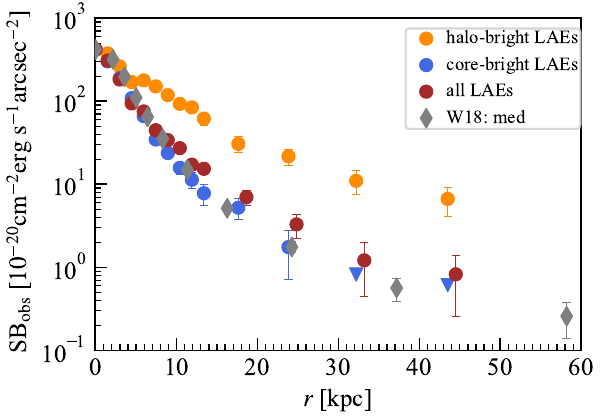}
    \caption{Ly$\alpha$ median-stacked radial surface brightness profiles of the LAEs in the two samples. For comparison, the grey diamonds show also the median profile of $3<z<4$ LAEs from \citet{Wisotzki_2018}. All SB profiles are rescaled to match at $r=0$. SB$_\mathrm{obs}$ $1\sigma$ upper limits are indicated by triangles. All SB values are re-scaled to $z=3.5$ by a factor of $[(1+z)/(1+3.5)]^4$. Small horizontal shifts between the different sets are applied for visualization purposes only. 
    }
    \label{Fig_flat_stack}
\end{figure}

Inspecting the radial SB profiles of our two sets of LAEs (Fig.~\ref{Fig_SB_profiles}) we find that overall, the halo components are very well described by single exponentials, with no systematic residuals suggesting that a different parametrisation might be preferable. Note that many of the core-bright LAEs have halo scale lengths of just a few kpc, implying that their overall profiles are significantly influenced by PSF blurring. This causes them to display some degree of outwards curvature, resulting in a mild trend to flatten at larger radii. This however is fully captured by the PSF-convolved exponential halo models. 

Only in a few halo-bright objects  (IDs 2726, 4842, 400, 8284) we see surface brightness profiles that beyond $\sim$15--20~kpc do not decline as rapidly as an exponential, often almost approaching constant values. A comparison with Fig.~\ref{Fig_NB_images} shows that this is always caused by secondary peaks of Ly$\alpha$ emission adding flux at larger distances. In particular, ID~2726 has a strongly lopsided LAH suggestive of a close companion, ID~4842 and ID~400 display several minor emission spots within and outside the plotted 50~kpc radius, and ID~8284 is evidently part of a more complex structure, possibly a merger of two LAEs. On the other hand, 5/9 of the halo-bright LAEs have profiles that -- while much more extended than the core-bright objects -- are fully consistent with the exponential fits.

Until now, SB measurements of these outer regions of LAHs around the low-mass galaxies have been performed almost exclusively in stacked profiles based on combining large numbers of objects \citep{Wisotzki_2018, Gallego_2021, Lujan_Niemeyer_2022, Kikuchihara_2022, Guo_2024_SB_profiles}, due to the extremely low SB levels required. Our individual LAH measurements reveal a large dispersion of halo sizes and profile shapes, which cannot be captured in stacked data. To compare with previous results and assess the representativity of our small sample, we also perform a stacking analysis by median-combining the radial profiles of the individual objects in the two subsamples. The result is presented in Fig.~\ref{Fig_flat_stack}, together with the stacked profile by \citet[][W18 hereafter]{Wisotzki_2018} from $\sim$100 faint LAEs in two MUSE deep fields. For display purposes and to enable the comparison of shapes, all stacked profiles are rescaled to match at the center.

We find that the core-bright stack is similar (within the error bars) to the W18 profile, whereas -- by design -- the halo-bright stacked profile reaches much higher SB levels in the outer regions. Stacking the union of both subsamples results in a combined profile only very slightly above the core-bright stack, underlining the robustness of the median against a small number of outliers.

\subsection{Ly$\alpha$ emission from nearby faint sources}
\label{discrete_neighbors}

Several studies have argued that the superposition of emission from faint LAEs in the surroundings of high-redshift galaxies should make a relevant contribution to the outer regions of observed Ly$\alpha$ SB profiles \citep[e.g.][]{Lake_2015, Mas_Ribas_2017, Mitchell_2021, Byrohl_2021}. Here we test and quantify this hypothesis using our set of individual ultradeep Ly$\alpha$ profiles.

\subsubsection{Expected mean contribution from undetected sources}
\label{discrete_neighbors_expected}

First, it is important to clearly define our terminology. Some authors use ``satellites'' when referring to faint nearby LAEs that are undetected as individual sources but contribute to the measured integrated Ly$\alpha$ SB \citep{Momose_2016, Mas_Ribas_2017}. This can be appropriate for relatively massive central galaxies, but typical faint LAEs residing in Dark Matter halos with masses of $\la 10^{11}\:M_\odot$ \citep[e.g.,][]{Khostovan_2019, Herrero_Alonso_2023A} have virial radii of only around $\sim$20--40~kpc, and Halo Occupation Distribution modeling indicates that the mean satellite fractions are very low \citep{Herrero_Alonso_2023A}. We therefore prefer to speak of (apparent) neighbors rather than satellites, explicitly including two halo term associations, i.e.\ large-scale structure (\citealt{Byrohl_2021} use the term ``other halos''). We also note that in a stack of photometric narrowband imaging data with a bandwidth of 50--100~\AA, the contributions from projected, physically unrelated fore- and background objects are not negligible \citep[see the discussion of bandwidth effects in][]{Herrero_Alonso_2023}.

For the present study it is furthermore important to realise that the impact of close neighbors on the observed Ly$\alpha$ morphology and the derived SB profile differ between a single system and a stack of many similar but randomly oriented galaxy-neighbor associations. When stacking, the combined contributions of all nearby objects will be smeared out over $2\pi$ in azimuth, also depending on the details of the stacking approach -- such as whether images or extracted profiles are combined, and whether the mean, the median, or some other estimator is applied. On the other hand, the imprint of neighbors onto the Ly$\alpha$ emission around a \emph{single} object depends strongly on the prominence of the neighboring source(s): A single bright neighbor makes the combined emission appear lopsided (as in ID~2726 of our sample), whereas a system surrounded by many discrete faint ``satellites'' on all sides would be indistinguishable from a diffuse halo (ID~4842 might be an example). 

The emissivity of genuine circumgalactic gas is expected to decline rapidly in the outwards direction. Far out at very low surface brightness levels it is therefore clear that the combined emission of neighboring sources will eventually exceed the intrinsic CGM emission. In other words, at some SB the radial profiles of individual LAHs with very different intrinsic properties (scale lengths, in particular) should converge towards a common level. In the following we compare our measurements with predictions of the level of this integrated external contribution $\mathrm{SB}_{\mathrm{unb}}$ to the radial SB profile, considering only those neighbors fainter than the formal detection limit for an LAE to be identified as a separate individual source. 

To estimate this contribution for a given survey with known flux limit, we have to make assumptions about the mean LAE number density at a given redshift (i.e.\ the Ly$\alpha$ luminosity function), in particular its faint-end shape, and about the enhancement of this number density due to clustering. Such a calculation was first performed by \citet{Herrero_Alonso_2023} who compared the outcome with the stacked Ly$\alpha$ SB profile by W18. Here we use the same formalism with updated luminosity function parameters (summarized for convenience in Appendix~\ref{ap:SB_unb_calculation}), but now apply it on an object-by-object basis to obtain predictions of the two halo term contribution to the integrated SB profile for each of our objects.

\begin{figure}
\centering
\includegraphics[width=\hsize]{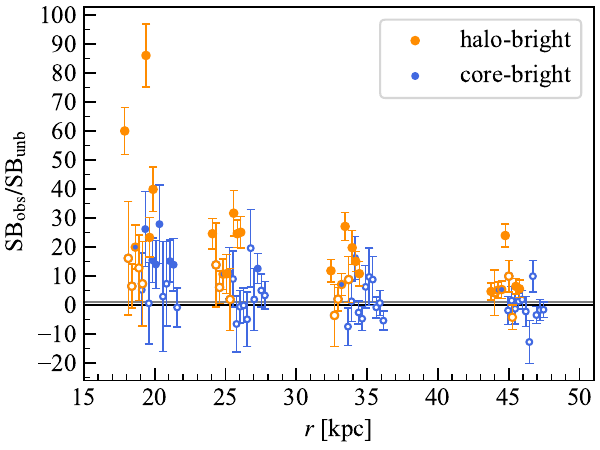}
\caption{Ratios of our measured SB profiles at $r > 15$~kpc to the predicted integrated Ly$\alpha$ surface brightness values SB$_\mathrm{unb}$ from undetected LAEs, calculated for each object separately as described in the text. Horizontal lines correspond to 0 and 1 (for the case of SB$_\mathrm{obs}$=SB$_\mathrm{unb}$) Small horizontal shifts are applied between the datapoints for better visualization, but each group of points belongs to the same fixed annulus defined in Sect~\ref{radial_profiles}. SB$_\mathrm{obs}$ values consistent with SB$_\mathrm{unb}$ within 1$\sigma$ uncertainties are presented as open circles.
}
\label{Fig_contribution_LAEs}
\end{figure}

In Fig.~\ref{Fig_contribution_LAEs} we compare our observed SB profiles with the result of this calculation, expressed as SB ratios $\mathrm{SB}_{\mathrm{obs}}/\mathrm{SB}_{\mathrm{unb}}$ to remove the $(1+z)^4$ dependence due to cosmological dimming. For the first radial bin at $15<r<20$~kpc we find that the measured SB values of most our objects are substantially above the expectation from undetected LAEs, $\mathrm{SB}_{\mathrm{obs}}/\mathrm{SB}_{\mathrm{unb}} \gg 1$, strongly indicating an internal (i.e. circumgalactic) origin of the emission especially in the halo-bright sample. Moving outward, the measured SB values decrease while $\text{SB}_{\text{unb}}$ is almost constant with radius, implying that the contribution expected from undetected neighbors is gaining in relevance. However, all of our significantly nonzero datapoints are well in excess of $\text{SB}_{\text{unb}}$, at all radii. Only in our outermost radial bin are most individual measurements consistent with $\text{SB}_{\text{unb}}$, although the error bars indicate that at these radii even our ultra-deep data are reaching their limits.

\subsubsection{The impact of different environments}
\label{discrete_neighbors_env}

\begin{table}
\caption{Overdensities, SFR and stellar masses of the LAEs.}  
\renewcommand{\arraystretch}{1.4}
\label{table:3}      
\centering          
\begin{tabular}{l l l l} 
\hline\hline       
ID & $\delta+1$ & $\log$ SFR & $\log$ M$_*$ \\ 
\hline
6700 & $4.8\pm1.8$ & $0.18^{+0.78}_{-0.43}$ & $9.1^{+0.4}_{-0.7}$ \\ 
106  &  $0.6\pm0.6$ & $0.39^{+0.56}_{-0.27}$ & $9.1^{+0.4}_{-0.4}$ \\ 
180  &  $1.3\pm0.9$ & $0.21^{+0.50}_{-0.25}$ & $9.1^{+0.3}_{-0.4}$ \\ 
6690  &  $1.8\pm1.2$ & $-0.66^{+0.65}_{-0.26}$ & $8.2^{+0.4}_{-0.5}$ \\  
6298  &  $2.2\pm1.2$ & $-0.32^{+0.41}_{-0.24}$ & $8.3^{+0.4}_{-0.4}$ \\   
469 &  $2.7\pm1.4$ & $-0.48^{+0.48}_{-0.27}$ & $8.4^{+0.3}_{-0.4}$ \\  
6929 &  $3.7\pm1.7$ & $-$ & $-$ \\ 
7187 &  $0.8\pm0.7$ & $-0.25^{+0.49}_{-0.20}$ & $8.6^{+0.2}_{-0.4}$ \\   
590 &  $0.9\pm0.8$ & $-1.01^{+0.52}_{-0.23}$ & $7.6^{+0.5}_{-0.4}$ \\ 
305 &  $1.2\pm0.9$ & $-0.61^{+0.53}_{-0.23}$ & $8.3^{+0.2}_{-0.3}$ \\ 
8327 & $4.2\pm1.8$ & $-0.08^{+0.50}_{-0.28}$ & $8.8^{+0.3}_{-0.3}$ \\ 
660 & $5.4\pm1.9$ & $-0.58^{+0.59}_{-0.31}$ & $8.3^{+0.4}_{-0.5}$ \\ 
2726 & $2.1\pm1.1$ & $-0.28^{+0.64}_{-0.34}$ & $8.6^{+0.5}_{-0.4}$ \\ 
4842 & $2.7\pm1.3$ & $-0.77^{+0.72}_{-0.29}$ & $8.2^{+0.3}_{-0.4}$ \\    
400 & $2.2\pm1.2$ & $-0.75^{+0.63}_{-0.31}$ & $8.1^{+0.3}_{-0.4}$ \\   
6297 & $1.7\pm1.1$ & $-0.33^{+0.44}_{-0.25}$ & $8.4^{+0.3}_{-0.3}$ \\ 
7586 & $3.1\pm1.4$ & $-0.86^{+0.53}_{-0.25}$ & $7.9^{+0.4}_{-0.4}$ \\ 
8360 & $3.4\pm1.4$ & $-1.21^{+0.67}_{-0.23}$ & $7.6^{+0.6}_{-0.3}$ \\ 
8537 & $5.3\pm1.9$ & $0.97^{+0.34}_{-0.28}$ & $9.8^{+0.2}_{-0.4}$ \\ 
8377 & $5.1\pm1.8$ & $0.55^{+0.52}_{-0.37}$ & $9.6^{+0.2}_{-0.3}$ \\ 
8284 & $4.1\pm1.8$ & $-$ & $-$ \\ 
     
\\ 
\end{tabular}
\tablefoot{$\log$~SFR is in M$_\odot/$yr, $\log$~M$_*$ is in M$_\odot$ (provided in \citealt{Bacon_2023})}
\end{table}

\begin{figure*}
    \centering
    \includegraphics[width=\hsize]{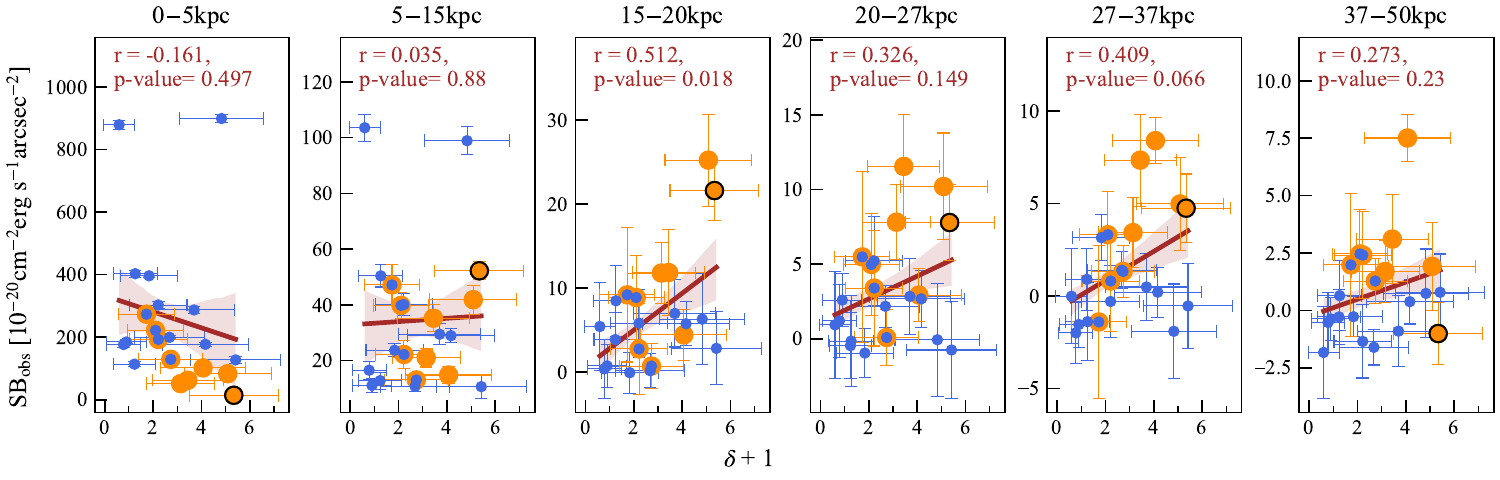}
    \caption{Observed Ly$\alpha$ surface brightness versus LAE environmental density $\delta+1$, estimated as described in the text. Each panel represents a different radial range, as indicated. The straight lines show the best-fit linear regression, the shaded regions the 68\% confidence interval). The numbers provide the strength of correlation characterized as as Pearson coefficient $\mathrm{r}$ and the probablility $p$ for the null hypothesis of no correlation. The special case of ID~8537 with its central Ly$\alpha$ depression is indicated by the black-edged symbol; it is excluded from the regression in the innermost radial bin. All SB values are re-scaled to $z=3.5$ by a factor of $[(1+z)/(1+3.5)]^4$. 
    }
    \label{Fig_SB_overdensity}
\end{figure*}

In the previous subsection we adopted a statistical correction for clustering, but so far we did not take into account that the expected number counts of surrounding faint neighbors should be affected by the environmental density of each object differently. We now investigate whether the Ly$\alpha$ SB in the outer halo regions correlates with the measured overdensities around the objects.

There are many ways of defining the concept of ``environment'' for galaxies. Here we choose a particularly simple approach based on the fact that the transverse field of view in our pencil-beam survey subtends only a few hundred (physical) kpc at our redshifts. This implies that any large-scale structure manifests itself mainly in $z$ direction, visible as strong modulations in the redshift histogram (see Fig.~1 in \citealt{Herrero_Alonso_2021} for an illustration). 

We assign an ``overdensity'' $\delta$ to each LAE in our sample based on the following prescription: We use the full AMUSED sample of LAEs in the Hubble Ultra-Deep Field to construct a smooth quasi-continuous redshift distribution by applying a KDE (Kernel Density Estimator) filter to the redshift catalog with a Gaussian Kernel of 600 km/s (FWHM), the same as adopted for the spectral bandwith by \citet{Herrero_Alonso_2023}. The ratio of observed and expected counts per redshift bin, each filtered by the same Kernel, gives the desired (over)density estimate, here expressed as the density ratio, $\delta + 1$. A Poissonian error bar for $\delta$ is estimated from the \emph{expected} number of sources at the location of each object. Note that this definition of density accounts also for spectral sensitivity variations of the MUSE instrument and for the masking effects of terrestrial night sky emission. Values of $\delta + 1$ for individual objects are presented in Table~\ref{table:3}.

In Fig.~\ref{Fig_SB_overdensity} we present the distribution of Ly$\alpha$ SB measurements (including detections consistent with zero) in relation to the corresponding environmental densities in the following radial bins: the innermost aperture ($r<5$~kpc) corresponds mostly to the host galaxy itself and the inner halo after blurring by the MUSE PSF; the second radial bin (5~kpc $<$ $r$ $<$ 15~kpc) contains most of the halo emission from around $\sim$1--2 exponential scale lengths, the remaining annuli gradually trace more and more of the outer halo regions. There are several interesting observations to be made in this sequence of plots:

Firstly, the central Ly$\alpha$ surface brightness of LAHs (i.e. in the inner halo region $r<5$~kpc and in an intermediate annulus at 5~kpc $<$ $r$ $<$ 15~kpc) is not significantly correlated with density, and an apparent weak anti-correlation at $0<r<5$~kpc is not statistically significant; it would most likely weaken by additional SB measurements and it practically disappears for alternative radial binning (e.g., $0<r<10$~kpc for the innermost bin). This demonstrates that the emission from the inner halo region is a purely internal phenomenon, driven by processes happening inside the galaxy and largely decoupled from the surrounding environment.

A rather different picture emerges in the outer halo. Our measurements suggest a positive correlation between density and Ly$\alpha$ SB at all radii $> 15$~kpc, although there is still substantial scatter and the correlation is formally significant at more than 5\% confidence only for the 15--20~kpc annulus. Nevertheless, the four rightmost panels resemble each other closely, despite the different SB levels involved, and taken together it seems clear that some environmental influence must be present. The similarity of the relation between SB and $\delta+1$ in all four radial bins strongly suggests a similar origin of the extended Ly$\alpha$ emission. 

Particularly remarkable in Fig.~\ref{Fig_SB_overdensity} is the stark change from $r<5$~kpc to $r>15$~kpc. And indeed the intermediate annulus at 5--15~kpc reveals that a turnover from purely internal to externally fed diffuse gaseous halos is happening within this range.

To conclude, our data show that the outer zones of Ly$\alpha$ halos are much more receptive to environmental influences than the inner halo regions. Within the range probed by our radial profiles we see however that a significant ``other halos'' contribution due to satellites or close neighbors is limited at most to the outermost radial bin for the halo-bright sample, while at lower radii the observed emission is a few times higher than expected contribution (see also Fig.~\ref{Fig_contribution_LAEs}).

Presumably contribution from SB$_\mathrm{unb}$ becomes dominant only at even larger radii and lower surface brightnesses. The observed Ly$\alpha$ emission in the outer regions of LAHs therefore might also originate in diffuse cool gas extending the CGM outwards into the IGM. However it remains unclear at this point how the Ly$\alpha$ emission from these regions is actually formed and what the main powering sources are. We investigate this question in the following subsection.

\subsection{Powering the Ly$\alpha$ emission in the outer halos}
\label{UV_ionisation}

We now discuss which radiative processes can account for the observed Ly$\alpha$ emission at large radii. We consider different contributions to the local UV radiation field from star formation in the host galaxy, the metagalactic UV background, and local enhancements in the radiation field due to overdensities in the galaxy distribution, including the possible influence of nearby active galactic nuclei (AGN).

\subsubsection{Local UV radiation field}
\label{UV_local}

\begin{figure*}
    \centering
    \includegraphics[width=\hsize]{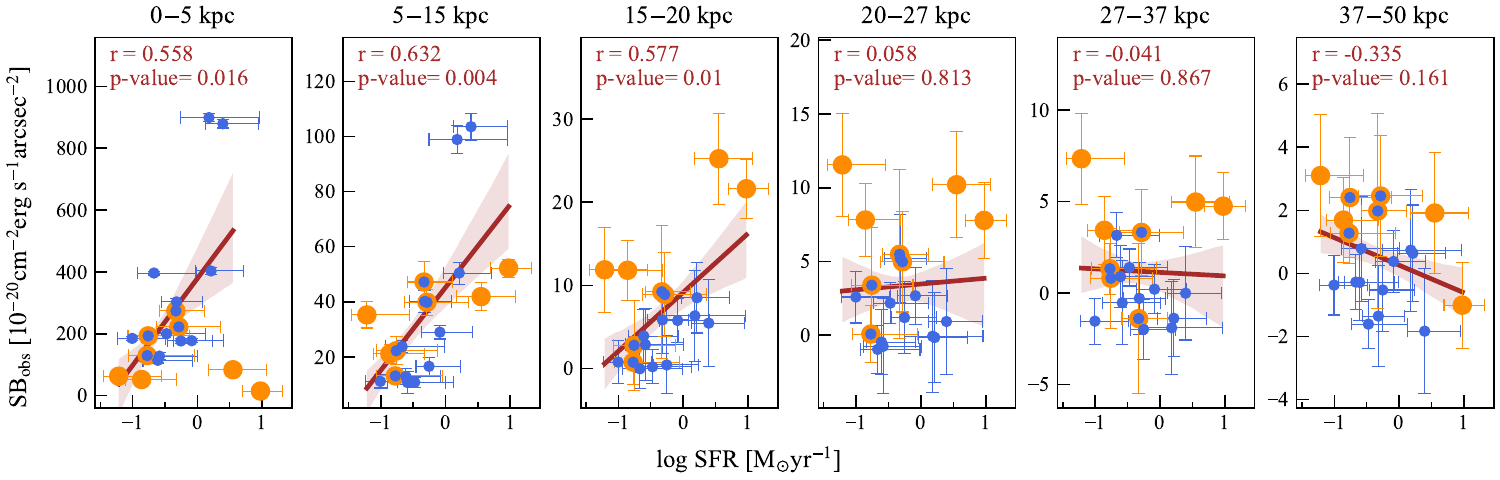}
    \caption{Observed Ly$\alpha$ surface brightness versus star formation rate (estimated as described in the text) for all LAEs in our sample with detected UV counterparts in deep HST imaging. As in Fig.~\ref{Fig_SB_overdensity}, the plots also show linear regression lines with Pearson coefficients and probabilities for the null hypothesis of no correlation.
    }
    \label{Fig_SB_vs_SFR}
\end{figure*}

Ionizing photons produced by massive stars from recent star formation inside the central galaxy are expected to be the dominant contribution to the total photoionization rate close to the galaxy. The production rate $R$ of Lyman continuum (LyC) photons can be estimated as roughly proportional to the star formation rate (SFR), with about $10^{53}$ photons per second for SFR = $1\:M_\sun$~yr$^{-1}$ \citep{Madau_1999}. 

The star formation rates of the galaxies in our sample are taken from the AMUSED catalog \citep[][; see Table~\ref{table:3}]{Bacon_2023}, based on modeling the broad-band photometric spectral energy distributions of the HST-detected counterparts to our LAEs. AMUSED provides SFR estimates derived with two different modeling codes, Magphys \citep{da_Cunha_2008} and Prospector \citep{Johnson_2021}, but we checked that the choice between these two has negligble impact on our findings. Here we adopt the results from Prospector, specifically the SFRs averaged over the last 100~Myr.  

Figure~\ref{Fig_SB_vs_SFR} presents the empirical relation between Ly$\alpha$ surface brightness and the SFR of the host galaxy for our two subsamples, separately for the same six annular apertures as in Fig.~\ref{Fig_SB_overdensity}.

 Our first finding from Fig.~\ref{Fig_SB_vs_SFR} is a very significant positive correlation between SFR and Ly$\alpha$ emission in the inner regions -- not just the central aperture but extending out to $\sim$20~kpc. This is of course expected if the Ly$\alpha$ photons were initially generated as recombination radiation after photoionization by massive stars. Essentially these panels contain a variation of the well-known relation between the UV absolute magnitude and the total Ly$\alpha$ luminosity of LAEs \citep{Santos_2021}, H$\alpha$-based SFR versus the total Ly$\alpha$ luminosity \citep{Hayes_2014} or UV-based versus Ly$\alpha$-derived SFRs \citep{Blanc_2011}, but now for the first time split up into distinct radially resolved Ly$\alpha$ regimes. Note that in Fig.~\ref{Fig_SB_vs_SFR} the $y$ axis shows the directly measured mean surface brightness, avoiding any possible hidden correlation of Ly$\alpha$ output with SFR through a common dependence on luminosity distance that influence other representations. 

Interestingly, the correlation between Ly$\alpha$ SB and SFR disappears entirely for $r \ga 20$~kpc. One might even imagine an anticorrelation forming in the outermost annulus, but this trend is speculative at best and certainly not statistically significant (also considering that the apparently negative SB values all have error bars that make them consistent with zero). But it is clear that there is a break around $\sim$20~kpc, i.e.\ at roughly 4--5 exponential scale lengths for typical Ly$\alpha$ halos \citep{Leclercq_2017}. The radiative processes in the outer halo regions must be quite different from those in the inner parts. 

This change in the correlation between SB$_{\mathrm{obs}}$ and SFR in the outer halo becomes even more remarkable if considered together with the inverse behaviour of the Ly$\alpha$ SB vs.\ environmental density relation revealed above (Sect.~\ref{discrete_neighbors_env}). While the turnover of that relation might occur at a somewhat smaller radius, the basic trends are clearly visible: The Ly$\alpha$ emission from the central regions and inner halos scales significantly with the SFR but not with the environment. The outer Ly$\alpha$ halo, on the other hand, responds mainly to the environment but not at all to the recent star formation in the host galaxy. Before we discuss the implications of these findings in Sect.~\ref{section_discussion} we now explore the contributions of external UV sources to the radiation field and the possible impact on the outer LAH emission.

\subsubsection{Fluorescence from the UV background}
\label{UV_UVB}

\begin{figure}
\centering
    \includegraphics[width=\hsize]{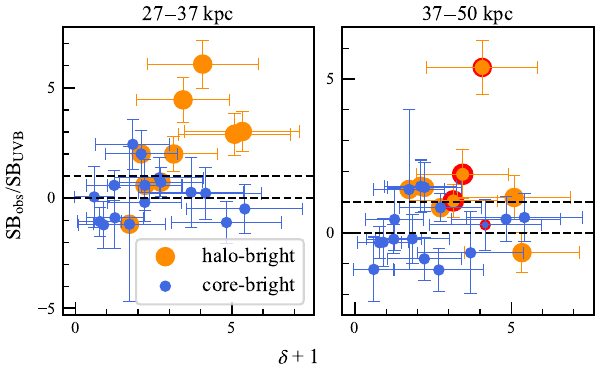}
    \caption{Maximum fractional contribution from UVB fluorescence to the Ly$\alpha$ emission at $27<r<37$~kpc (left) and $37<r<50$~kpc (right). Each panel shows the ratio of the observed SB to the predicted SB (upper limit assuming $f_{\text{c}}=1$, see text), $\mathrm{SB}_{\mathrm{obs}}/\mathrm{SB}_{\mathrm{UVB}}$, plotted versus environmental density. The horizontal dashed black lines represent constant values of 0 and 1, respectively. Objects with nearby luminous AGNs are marked with red edges: thin edge correspond to ionization rate enhancement compared to the UVB of $\sim 2$, thick edges of $\sim 10$.
    }
\label{Fig_contribution_UVB}
\end{figure}

A natural way to enhance Ly$\alpha$ emission in the outer regions is recombination in response to Lyman continuum (LyC) photons from the global UVB impinging on layers of neutral hydrogen in the CGM. This process -- sometimes referred to as ``UVB fluorescence'' -- has been discussed repeatedly in the literature  \citep[e.g.,][]{Cantalupo_2005, Furlanetto_2005, Kollmeier_2010, Gallego_2021}. The resulting Ly$\alpha$ SB of a single \ion{H}{i} cloud optically thick to LyC radiation is proportional to the metagalactic hydrogen photoionisation rate $\Gamma_{\ion{H}{i}}(z)$, modulated by cosmological surface brightness dimming. For an ensemble of clouds an additional uncertainty comes from the poorly known covering factor $f_{\text{c}}$ of circumgalactic \ion{H}{i} exposed to the UVB. To estimate the surface brightness, we follow \citet{Cantalupo_2005} and \citet{Gallego_2021}, but using an updated model of the UVB taken from \citet{Puchwein_2019}. In Appendix~\ref{ap:uvb_calculation} we summarize the core assumptions and relations. Note that again we perform this calculation separately for the redshift of each object in our sample rather than for an averaged stack as in \citet{Gallego_2021}. 

For a cloud covering fraction of $f_{\text{c}}=1$ (single cloud approximation) we obtain an expected surface brightness of the order of $10^{-20}$ erg s$^{-1}$ cm$^{-2}$ arcsec$^{-2}$ for the redshift range of our sample, in good agreement with previous estimates. While for the CGM of real galaxies the covering fraction is probably much below unity (\citealt{Gallego_2021} estimate a mean  $f_{\text{c}} \simeq 0.2-0.3$ for the our redshifts and radial range), here we adopt $f_{\text{c}}=1$ and consider the results as an upper limit to the predicted emission. 

In Fig.~\ref{Fig_contribution_UVB} we present the results of our object-by-object application of Eq.~\eqref{Eq_SB_uvb} as ratios of observed to predicted SB, a rescaling which removes the otherwise strong influence of cosmological $(1+z)^4$ dimming. A ratio of $\mathrm{SB}_{\mathrm{obs}}/\mathrm{SB}_{\mathrm{UVB}} = 1$ thus implies that all of the observed emission could in principle be powered by the UVB if the maximal possible covering fraction of unity is assumed (and correspondingly less for $f_{\text{c}}<1$). 

Figure~\ref{Fig_contribution_UVB} shows that in the second-outermost radial bin (27--37~kpc) there are several points well above the expected emission level of a UVB-dominated LAH, especially if $f_{\text{c}}\simeq 0.3$  rather than 1. This follows from the fact that setting $f_c = 0.3$ reduces $\mathrm{SB}_{\mathrm{obs}}/\mathrm{SB}_{\mathrm{UVB}}$ by a factor of $0.3$. For those objects the dominant emission mechanism at these radii cannot be UVB fluorescence. On the other hand, several of our LAHs, especially among the core-bright sample, have SB values at this radius consistent with the UVB prediction, but also consistent with zero given the error bars. We have to acknowledge that our individual object measurements are still not sensitive enough to strongly constrain the presence of UVB fluorescence in the outer halos. This limitation is even more apparent in the 37--50~kpc range, shown in the right-hand panel of Fig.~\ref{Fig_contribution_UVB}, where only one object has a $\mathrm{SB}_{\mathrm{obs}}/\mathrm{SB}_{\mathrm{UVB}}$ ratio significantly greater than 1.

\subsubsection{Ly$\alpha$ emission due to a locally enhanced radiation field?}
\label{UV_enhanced}

At $3<z<4$ the IGM is quite transparent for EUV radiation, with a mean free path for LyC photons of around 40--100~Mpc \citep{Worseck_2014}. While the UVB should therefore be quite uniform on larger scales, it can still be locally boosted by an overabundance of strong UV sources such as luminous star forming galaxies or AGN. An interesting quantity in this context is the distance from a galaxy where the local photoionization rate becomes equal to that due to the metagalactic UVB (see Appendix~\ref{ap:uvb_calculation}). For the objects in our sample we obtain $r_{\text{eq}}$ values between $\sim$1 and 10~kpc for $f_{\text{esc,LyC}}=0.01$ and between $\sim 4$ and 40~kpc for $f_{\text{esc,LyC}}=0.08$ \citep{Chisholm_2018}, with the LyC escape fraction $f_{\text{esc,LyC}}$ as the main free parameter.

Since none of the objects in our sample has a bright neighboring LAE at such close distance, any possible enhancement -- if present -- could come only from a general overdensity of LAEs in the environment, or from a very luminous nearby AGN. We first consider the evidence of LAH emission potentially boosted by the LyC radiation from an LAE overdensity. This is evidently a question closely related to the one asked in Sect.~\ref{discrete_neighbors_env}. Figure~\ref{Fig_contribution_UVB} therefore shows the surface brightness ratios $\mathrm{SB}_{\mathrm{obs}}/\mathrm{SB}_{\mathrm{UVB}}$ once more plotted against the environmental density ratios $\delta+1$. The left-hand panel shows a trend very similar to what is seen in Fig.~\ref{Fig_SB_overdensity}, in that several of the halo-bright objects correlate weakly with environment. However, this trend vanishes in the right panel, i.e.\ in the outermost radial bin. Apart from a single object (ID~8284) there is no indication that higher SB values are found in more overdense regions. Since the effects of external fluorescence should be most prominent at the largest radii (unless the covering fraction decreases steeply), we conclude that we do not see any evidence of a boosted local UVB due to LAE clustering effects.

Another possible origin of a locally boosted radiation field could be additional LyC photons from one (or several) nearby AGN. We therefore searched whether any our objects might be located within  the proximity zones of AGN detected in the 7~Ms X-ray survey of the ``Chandra Deep Field South'' \citep{Luo_2017}. Indeed, we find that 4 objects in our sample belong to two large-scale overdensities that each contain a known luminous AGN (further CID stands for IDs from``Chandra Deep Field South'' in \citealt{Luo_2017}). (1) ID~7586 and ID~8360, both at $z=3.07$, are 790 and 807~kpc from the type~2 AGN ID~1056 (in the MUSE-MOSAIC), or CID~746. This object is surrounded by a prominent, but not unusually luminous Ly$\alpha$ nebula \citep{den_Brok_2020} and belongs to a spectacular Ly$\alpha$ filament of more than 1~Mpc (proper) length that includes our two LAEs \citep{Bacon_2021}. (2) ID~8284 and ID~8327, both at $z=3.7$, are located 1.95 and 2.09 Mpc from the extremely luminous type-2 QSO CID~551 (ID~115003085 in the MUSE-Wide) discovered by \citealt{Norman_2002} (referred to as CDF~202). This AGN is also associated with a giant Ly$\alpha$ nebula \citep{den_Brok_2020}.

We estimated the radiative impact of these two AGN at the locations of our LAEs based on the assumption that the measured and absorption-corrected X-ray luminosities (taken from \citealt{Norman_2002}) can be converted to predict their LyC output assuming isotropy (a very generous assumption given that both objects are observed to be type-2 AGN). We follow the calculation in \citet{Gallego_2021} and obtain enhancement factors $\Gamma_{\ion{H}{i}, \text{ AGN}} / \Gamma_{\ion{H}{i}, \text{ UVB}}$ of the hydrogen photoionization rate over the metagalactic UVB of $\sim$10 for our two objects at $z=3.07$ and of $\sim$2 for the two LAEs at $z\simeq 3.7$. While these are certainly not insignificant values, we do not see any evidence of the measured Ly$\alpha$ surface brightnesses of the four LAEs in question to correlate with the estimated enhancement due to AGN. This is demonstrated in Fig.~\ref{Fig_contribution_UVB} where the four objects in AGN proximity zones are specifically marked with red edges. Only for the exceptionally high datapoint belonging to ID~8284 there could be a substantial boosting effect by a factor $\sim$2; on the other hand, this object has the most complex small-scale environment of all of our objects (see Fig.~\ref{Fig_NB_images}), and there may well be other reasons for its bright outer LAH without invoking AGN influence. In any case none of the other three proximity-zone LAEs stand out in Fig.~\ref{Fig_contribution_UVB}. The calculated above enhancement factors of 2 and 10 are most likely massive overestimates because some level of UV obscuration (which made the AGN appear as type-2 objects) also applies in transverse direction.

We can not categorically exclude the possibility of presence of an undetected dusty star-forming galaxy that is located near an LAE and contributes additional ionizing photons. However, we consider this scenario unlikely to provide a common explanation for the observed extended Ly$\alpha$ emission. Sub-millimeter galaxies are rare massive systems \citep{Casey_2014} with number densities at least two orders of magnitude lower than those of typical LAEs \citep[e.g.,][]{Dudzeviciute_2020}. Moreover, in the whole HUDF there is only one such galaxy detected with extremely deep ALMA observations at $3<z<4$, and it is not at the redshift of any of our LAEs \citep{Walter_2016, Boogaard_2019}.

\section{Discussion} 
\label{section_discussion} 

\begin{figure*}
\centering
    \includegraphics[width=\hsize]{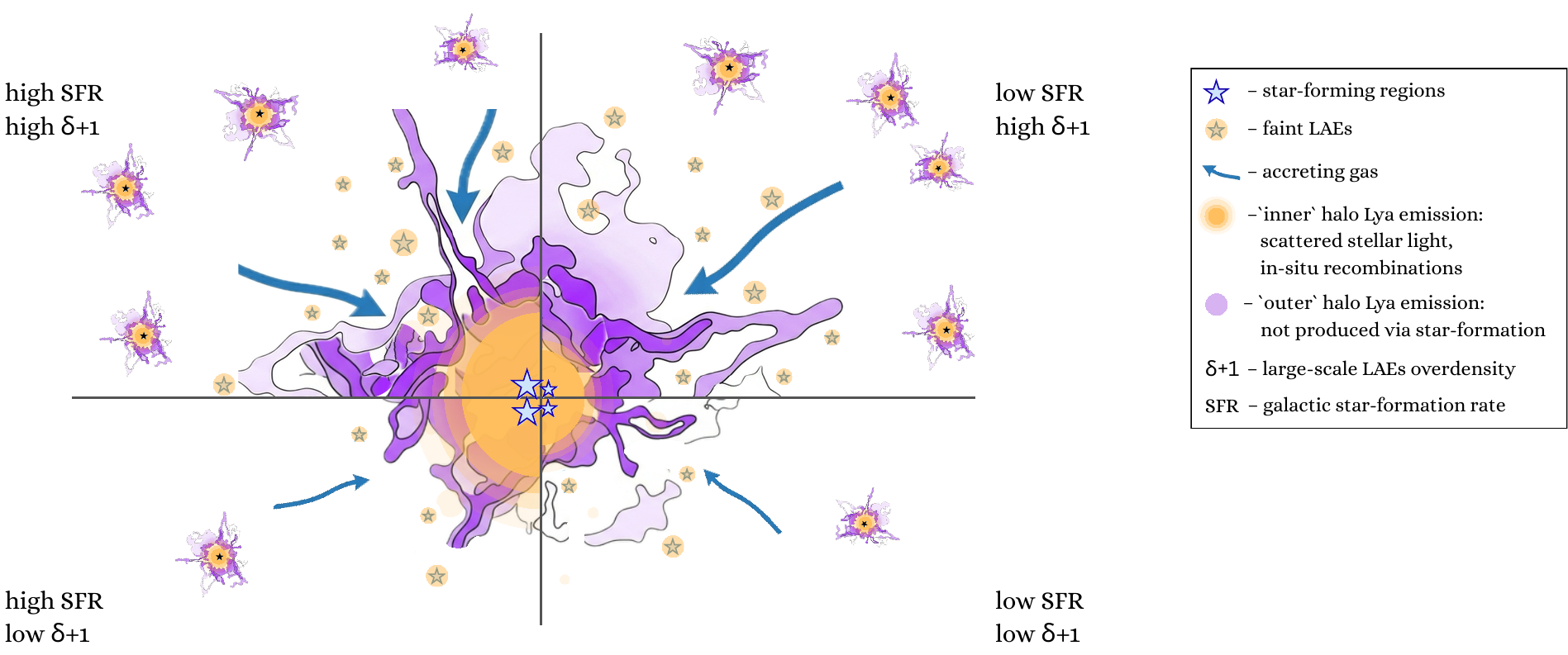}
    \caption{Artistic view of the different domains of LAHs. Four quarters picture four possible combinations of low/high SFR and $\delta+1$ that then manifest themselves via fainter/brighter inner (orange) and outer (violet) CGM Ly$\alpha$ emission. Blue stars refer to star-forming regions within galaxies, where larger stars indicate higher SFR. The large-scale overdensity is represented by the small orange-violet halos, as well as by a number of faint, individually unresolved LAEs (orange circles with a single star). The blue arrows represent accreting neutral gas from the IGM. 
    }
\label{fig:summary_sketch}
\end{figure*}

Our analysis of Ly$\alpha$ surface brightness profiles measured in the MUSE Extremely Deep Field has revealed remarkably clear differences between the inner and outer regions of Ly$\alpha$ halos. At radii up to $\sim$15--20~kpc the emission is closely correlated with the star formation rate in the central galaxy, but not at all with the environmental density (as derived from the number of nearby LAEs at the same redshift). At larger distances the scenario reverses: The correlation with central star formation rate disappears, while LAEs in richer environments show enhanced ``outer halo'' emission. These changes suggest a fundamental switch of the relevant emission mechanisms from internal powering to external regulation, as discussed in the following. The same analysis on core-bright sample only reveals the same trends, while halo-bright sample is only reinforcing them, especially at $r>20$~kpc. {We confirm that these results are robust against different radial binning.

\subsection{Inner Ly$\alpha$ halos are powered only by star formation}
\label{section_discussion_inner} 

A relation between the total Ly$\alpha$ luminosity $L_{\mathrm{Ly}\alpha}$ and the SFR of a galaxy is of course well established, for instance, through a connection with H$\alpha$ luminosity, that connects it to SFR \citep{Kennicutt_1998}. This method has been used many times to predict SFR from $L_{\mathrm{Ly}\alpha}$, with the often unknown Ly$\alpha$ escape fraction as the main parameter responsible for the substantial scatter in this relation. The results for our LAEs sample, consistent with overall typically observed LAEs in stellar mass, exponential halo scalelengths and halo fractions \citep[e.g., ][]{Wisotzki_2016, Leclercq_2017} show (Fig.~\ref{Fig_SB_vs_SFR}) that such a relation applies not only to the total Ly$\alpha$ emission (which is usually dominated by the bright central region close to the newly formed stars), but that it also holds for the extended halo at distances $r\ga 10$~kpc to the galaxy, corresponding to about 2--4 times the exponential scale length of the halo (and thus to about 20--40 times the scale length of the stellar body, cf.\ \citealt{Leclercq_2017}). Even at 15--20~kpc this correlation still seems to hold, although with more scatter; yet it is apparently gone in the subsequent radial bin at 20--27~kpc and beyond. For our sample we therefore tentatively locate a breakpoint of the dominant emission mechanism at about 20~kpc.

Interestingly, this is very close to the radius at which we find that environmental influence starts to become important (Fig.~\ref{Fig_SB_overdensity}). For $r<15$~kpc there is no evidence of any such influence on the measured halo brightness. The question of whether Ly$\alpha$ halos are different in overdense regions has been a topic of some interest itself, with somewhat controversial results so far. Note that previous authors only considered the halo scale lengths obtained by fitting exponential models to the data, finding either a significant increase in $r_\mathrm{s, h}$ with overdensity $\delta$ \citep{Matsuda_2012}, or no such dependence at all \citep{Xue_2017}, or no trend except for very large overdensities \citep{Kikuta_2023}. To compare with these studies, we also investigated the halo scale lengths and halo flux fractions of our LAEs, both of which are fully set by the bright inner halo regions (see Sect.~\ref{galfit_modeling}). Neither $r_\mathrm{s, h}$ nor ${F_\mathrm{h}/F_\mathrm{tot}}$ show any discernible correlation with environmental density.

It is informative to relate this transition radius to the approximate sizes $r_{\mathrm{vir}}$ of the dark matter halos in which these galaxies reside. Such estimates are very uncertain as they are based on global scaling relations. We estimate virial radii for our sample objects in three different ways: (i) by using the stellar masses derived from our SED fits (see Sect.~\ref{UV_local}) and applying the $M\star/M_{\mathrm{vir}}$ relation from \citet{Legrand_2019}; (ii) converting the Ly$\alpha$ luminosities into clustering-inferred virial masses using Fig.~8 of \citet{Herrero_Alonso_2023}; (iii) by applying the \citet{Kravtsov_2013} relation, updated for high redshifts by \citealt{Shibuya_2015} with resulting conversion factor of $0.02\pm0.01$ between stellar half-light radius and virial radius. Method (i) gives a mean value of $r_{\mathrm{vir}}\simeq19$~kpc given the mean stellar mass of $M\star\simeq10^{8.4}\mathrm{M}_\odot$ (Table~\ref{table:3}); method (ii) results in $r_{\mathrm{vir}}\simeq20$~kpc. A conversion of the measured $r_{s,c}$ (in Table~\ref{table:2} from F775W HST images) into half-light radii using factor of 1.68, applicable for the exponential disks, results in mean $r_{\mathrm{vir}}\simeq23$~kpc. 
Thus all three methods result in similar values of $r_{\mathrm{vir}}$ around $\sim$20~kpc, underlining that our sample consists of low-mass galaxies.

The transition radius of $\sim$15--20~kpc that we identified between ``inner'' and ``outer'' Ly$\alpha$ halo behaviour of our objects is thus very close to the typical virial radius. We therefore conclude that out to about $1 r_{\mathrm{vir}}$ the observed circumgalactic Ly$\alpha$ emission is mostly powered by massive stars, either as recombination radiation in the \ion{H}{ii} regions in the galaxy that is then scattered outwards by neutral gas in the CGM \citep{Zheng_2011, Kusakabe_2019}, or by direct recombination following in-situ photoionization due to Lyman continuum photons leaking into the CGM \citep[e.g. ][]{Mitchell_2021}. Accordingly, circumgalactic outflows driven by massive stars and supernovae presumably play the dominant role in shaping the distribution and physical state of the gas, largely indepenent of the larger-scale environment in which the galaxy resides. But the dominance of these internal power sources terminates roughly at the virial radius, beyond which environmental effects take over. Based on the same parent sample, \citealt{Guo_2024_SB_profiles} presented stacked Ly$\alpha$ profiles and reported a similar transition radius at $1 r_{\mathrm{vir}}$, beyond which the average surface-brightness profile deviates from an exponential form and begins to flatten.

\subsection{Outer Ly$\alpha$ halos are shaped by the environment}
\label{section_discussion_outer} 

Several mechanisms could contribute to the observed Ly$\alpha$ emission outside of (approximately) one virial radius, some of which we already evaluated quantitatively in Sect.~\ref{section_results}. We now weight the evidence for and against the different options.

We first consider the impact of neighboring ultrafaint and individually undetected LAEs at the same redshift. The fractional contribution of this impact is expected to increase with radius, and its absolute value should be higher in overdense regions. While Fig.~\ref{Fig_SB_overdensity} indicates that the SB in the outer halo indeed correlates with environmental density $\delta+1$, Fig.~\ref{Fig_contribution_LAEs} shows that this correlation holds also at radii where our measured SB levels are much higher than the predicted integrated SB due to the external LAEs. Only in our  outermost radial bin (37--50~kpc) are the individual measurements broadly consistent with the prediction, suggesting that here we might indeed see the onset of ``other halos'' (in the terminology of \citealt{Byrohl_2021}) dominating the Ly$\alpha$ SB profiles towards even larger radii.

The main uncertainty with this conclusion is the still poorly constrained faint-end slope of the Ly$\alpha$ luminosity function. As the baseline value, we choose $\alpha=-1.68$ (\citealt{Tornotti_2025}), which in our opinion represents the best faint-end constraints to date by taking advantage of the numerous MUSE surveys, including the deepest with $\mathrm{T}_{\mathrm{exp}}>90^{\mathrm{h}}$. This value is also consistent with the recent measurements by other groups (e.g. $\alpha=-1.63^{+0.17}_{-0.16}$ \citealt{Sobral_2018}, $\alpha=-1.58^{+0.11}_{-0.11}$ \citealt{de_La_Vieuville_2019}), which agree on a relatively shallow Ly$\alpha$ LF faint-end slope. Furthermore, $\alpha=-1.68$ is in good agreement with the preliminary results obtained by Pharo et al. (in prep.) on a comprehensive Ly$\alpha$ LF that includes the deepest MUSE observations of $\simeq2500$ LAEs at  $3<z<4$ and carefully accounts for selection effects. While the steeper faint-end of the Ly$\alpha$ LF would naturally increase the relative contribution $\mathrm{SB}_{\mathrm{obs}}/\mathrm{SB}_{\mathrm{unb}}$ (e.g., by up to a factor of $\simeq3$ for $\alpha=-1.84$ from \citealt{Herenz_2019}), but we consider this scenario unlikely given the recent results on large LAEs samples with better representation of the low-luminosity objects.

The rightmost panel of Fig.~\ref{Fig_SB_overdensity} shows one exceptionally high datapoint, which is the object ID~8284 that was already discussed briefly in Sect.~\ref{UV_enhanced}. Inspecting its Ly$\alpha$ NB image (Fig.~\ref{Fig_NB_images}), this object displays the most asymmetric LAH in our entire sample, possibly caused by the presence of a secondary LAE at $\sim$40~kpc separation just below the individual detection threshold of the MXDF survey. While the statistical rate of such relatively bright neighbors is much less than one per annulus (see Appendix~\ref{ap:SB_unb_calculation}), a chance occurrence is of course always possible. ID~106 might be another, although less extreme, object in our sample where the Ly$\alpha$ emission at the largest radii is boosted by a single faint neighbour. 

Nevertheless, the positive correlation of Ly$\alpha$ SB with environment at all radii $r>20$~kpc and therefore at different SB levels (see Fig.~\ref{Fig_SB_overdensity}) argues against discrete neighbors, which contribution does not differ much with radii,  as the main contributor to the outer halo. It is more likely that the observed Ly$\alpha$ emission at large radii originates in diffuse cool gas extending the CGM beyond the virial radius and possibly into the IGM. Richer environments would then naturally contain larger quantities of cool gas capable of shining in Ly$\alpha$. The remaining question would then be, what could be a plausible powering mechanism for this radiation, considering also the absence of a correlation with the central SFR.

Ly$\alpha$ photons can also be produced through collisional excitation of hydrogen followed by radiative de-excitation, e.g. emerging as cooling radiation from accretion flows into the CGM \citep[e.g. ][]{Haiman_2000, Furlanetto_2005, Faucher_Giguere_2010,Rosdahl_2012}. Ly$\alpha$ emission from cooling flows is expected to be particularly strong in massive halos \citep[e.g. ][]{Fardal_2001,Dijkstra_Loeb_2009,Ao_2020}, but according to high-resolution hydrodynamical simulations could still contribute significantly to the total Ly$\alpha$ budget also of low-mass systems \citep{Mitchell_2021}. Following these models, Ly$\alpha$ cooling radiation would be relevant mostly at radii large enough so that the central component becomes subdominant. We suggest that cooling radiation could provide a viable explanation for at least a substantial fraction of the observed outer halo emission. Since the inflow rate would be expected to scale with the environmental density but not directly with the central star formation rate, this would also provide a natural explanation for the reversal of correlation trends around $1 r_{\mathrm{vir}}$ found in our study.

Are there other possibilities to produce outer halo emission, for example fossil radiation after pa star formation event? This does not seem plausible since the SFR values used in our study are averaged over the past $10^8$~yr (Sect.~\ref{UV_local}), considerably longer than the hydrogen recombination timescale of $10^{4}/n_{\mathrm{e}}$~yr for the expected densities $n_{\mathrm{e}}\sim 10^{-4} - 10^{-2}$ \citep[e.g. ][]{Werk_2014}. While some recent studies reconstructing the star formation histories of LAEs \citep{Rosani_2020, Firestone_2025} have suggested that a substantial fraction may have experienced a major star-formation episode more than $200$~Myr prior to the current one, there should be no observable Ly$\alpha$ emission left from those events.

To summarize, we find that the correlation (or absence thereof) of Ly$\alpha$ surface brightness at a given radius with (i) central star formation rate and (ii) environmental density leads to distinctly different behavior at radii smaller or greater than (approximately) one virial radius. Figure~\ref{fig:summary_sketch} attempts to combine the different domains of Ly$\alpha$ halos into a single sketch. There, orange and violet color gradients represent inner and outer CGM emission, respectively, together with the sources of their regulation, so SFR (blue stars) and $\delta+1$ (other detected halos in orange-violet and undetected halos as orange circles with the single star). We also decode additional relevant information, which is not directly inferred from our results: (a) shades of violet represent an anisotropic \ion{H}{i} column density distribution (e.g. at lower $z$ traced by absorption lines in \citealt{Dutta_2026} or in the cosmological hydrodynamical simulation EAGLE in \citealt{Rahmati_2015}) and (b) the overall shape of the gas represents the varying \ion{H}{i} covering fraction \citep[e.g. see Fig.~10 in][]{Gallego_2021}, and (c) blue arrows represent neutral gas cold accretion from the IGM, a physical mechanism expected to feed galactic star-formation at $z\simeq3$ \citep[e.g. ][]{Keres_2009, van_de_Voort_2011}; such external accretion is naturally more prominent in regions with larger gas reservoirs, which are observationally traced by the LAEs overdensities \citep[e.g. ][]{Mukae_2020, Galbiati_2024}. Note that the undetected halos and accretion flows do not spatially coincide with the extended emission shown in our sketch. Rather, they illustrate, that their presence would contribute to and enhance the observed extended emission (in violet).

\section{Conclusions} \label{section_conclusions}

In this work we obtained new insights into the balance of different powering mechanisms for the extended Ly$\alpha$ halos surrounding most star-forming galaxies at high redshifts. While there is broad agreement that the inner regions of Ly$\alpha$ halos probe the cool gas in the circumgalactic medium, it has frequently been proposed that what appears as outer halo emission could in fact be a superposition of discrete neighboring LAEs (or ``satellites''). Previous investigations of this hypothesis were, however, all based on stacking and large samples averages, enforced by the low surface brightnesses involved, and therefore insensitive to trends and correlations between different observables. Here we measure, for the first time, the properties of individual Ly$\alpha$ halos at $3<z<4$ beyond their virial radii. The core-bright LAEs in our study is a representation of a general LAEs population with no criterion on halo brightness, while halo-bright LAEs reinforce the observed trends in outer halo.

The key results of our study are presented in Fig.~\ref{Fig_SB_overdensity} and Fig.~\ref{Fig_SB_vs_SFR} and could be summarized as follows:
\begin{enumerate}
\item While our sample consists of the brightest LAEs in the MXDF in terms of core flux and extended surface brightness, their Ly$\alpha$ luminosities are mostly below $10^{42}$~erg s$^{-1}$ and thus intrinsically even fainter than the samples used in most previous stacking studies (with the exception of \citealt{Guo_2024_SB_profiles}, which was also based on the MXDF). We estimate virial radii from three different observed quantities and methods, and obtain consistent mean values of $r_{\mathrm{vir}}\sim 20$~kpc for the sample. Our measurements out to distances of 50~kpc thus reach typically $2.5 r_{\mathrm{vir}}$.

\item In most objects the Ly$\alpha$ SB profile is well approximated by a single exponential out to the largest radii where emission is detected, although with a large dispersion of scale lengths and thus halo sizes (Fig.~\ref{Fig_SB_profiles}). In the few cases with significant deviations from a single exponential, these derivarions can be traced directly to secondary peaks of Ly$\alpha$ emission (Fig.~\ref{Fig_NB_images}). Stacking our small sample results in a median SB profile that is in excellent agreement with previous work (Fig.~\ref{Fig_flat_stack}).

\item The Ly$\alpha$ emission out to about one virial radius is tightly correlated with the SFR of the central galaxy (Fig.~\ref{Fig_SB_vs_SFR}), with SFRs defined as averages over the last 100~Myr based on SED modelling of the stellar broad band continuum. Such a  correlation is expected, but is here demonstrated for the first time to hold even at distances of 10--15~kpc (several halo exponential scale lengths) from the central galaxy. However, beyond 20~kpc, i.e.\ for $r\ga 1r_{\mathrm{vir}}$, this correlation is lost entirely, implying that the outer halo emission does not anymore respond to the UV radiation from centrally located young stellar populations.

\item Conversely, we find that the richness of the Mpc-scale environment of our LAEs only correlates with Ly$\alpha$ surface brightness in the outer halo regions (Fig.~\ref{Fig_SB_overdensity}) with a turnover at about 15--20~kpc, thus again close to $1 r_{\mathrm{vir}}$. The absence of any correlation with the environmental density at smaller radii strongly supports the interpretation of the inner Ly$\alpha$ halos as a genuine circumgalactic phenomenon mostly driven by internal processes such as outflows and gas recycling.

\item The correlation between Ly$\alpha$ SB and environmental density $\delta+1$ for $r \ga 1 r_{\mathrm{vir}}$ is not strong, but looks very similar in all of our radial bins (right panels of Fig.~\ref{Fig_SB_overdensity}), lending support to its statistical significance.  We interpret this trend, together with the absence of any significant correlation with SFR at these radii, as evidence that the outer regions of Ly$\alpha$ halos are indeed dominated by external factors. 

\item The contribution of individually undetected faint LAEs to the integrated SB at these radii is, however, insufficient to generate the observed trend in our data. While this contribution is fundamentally expected to become relevant at sufficiently large radii, for the radial range probed by our data its amplitude is still too small to have a measurable impact. 

\item Searching for ways to enhance the Ly$\alpha$ emission from outer halos, we consider fluorescence by the cosmic UV background. We find that this mechanism could provide a surplus to the measured SB at $37< r < 50$~kpc of up to $\sim 1\sigma$ where $\sigma$ is the typical error bar of our datapoints  (Fig.~\ref{Fig_contribution_UVB}). However, this is insufficient to explain the correlation between SB and environment.

\item We also investigate the possibility that some of our LAEs, especially in large overdensities, might receive a boosted UV radiation field by  illumination from luminous AGN in the vicinity. We find that while four objects might indeed experience significant boosting factors between 2 and 10, the only one of them showing enhanced outer halo emission also features a likely close companion and is therefore a questionable case of AGN boosting.

\end{enumerate}

We conclude that inner and outer Ly$\alpha$ halos constitute two distinctly different domains with a rather well-defined turnover at $\sim 1 r_{\mathrm{vir}}$. Note, however, that what we here call the ``inner'' region corresponds everything inside several halo exponential scale lengths and thus typically encompasses more than 90\% of the total Ly$\alpha$ flux from the halo. The outer halo, on the other hand, constitutes the gradual transition zone into the intergalactic medium, presumably with no well-defined outer boundary (other than observational sensitivity limits).

We have considered different options to explain the (extremely weak) observed Ly$\alpha$ emission from these outer halos through external influences such as UV fluorescence or the integrated contribution from faint neighboring LAEs. Since none of them seems to be sufficient to drive the observed correlation between emission and environment, the question arises: What then could be the origin of this emission? At this point it is worth recalling that most theoretical models and simulations of Ly$\alpha$ halos predict that collisional excitation, such as occurring in cold accretion streams \citep{Fardal_2001, Furlanetto_2005, Rosdahl_2012, Mitchell_2021, Blaizot_2023}, can contribute substantially to the Ly$\alpha$ output especially at large radii. The relevance of this process is very difficult to test empirically due to the dominance of other radiation mechanisms in the total emission. Here we speculate that the environmental influence on the outer halo properties might be explained by the accretion of cold gas from the surrounding medium. 

While the evidence presented in this paper is clearly insufficient to further elaborate and test this hypothesis, we see a clear route to do so in the future.  As key observable we identify the radial dependence of the Ly$\alpha$ spectral line profile. If the emission mechanisms are indeed fundamentally different between inner and outer halos, including different gas kinematics (e.g., outflow- vs.\ inflow-dominated), then this should get reflected in characteristic differences in the corresponding spectra. Such changes may be very challenging to detect, but not impossible: Recently, \citet{Gallego_2021} and \citet{Guo_2024_spectra} found from radially resolved spectral stacking that the Ly$\alpha$ emission peak at large radii tends to be significantly blueshifted relative to the central emission -- which in turn is usually redshifted with respect to systemic, presumably due to outflows \citep{Verhamme_2006, Blaizot_2023}. Of course these are average trends that could be hiding a large underlying diversity of processes. Thus, investigating the radial evolution of Ly~$\alpha$ line profiles across individual galaxies is a promising research avenue, that can potentially provide further evidence on the regime transition at $\sim 1 r_{\mathrm{vir}}$. Furthermore, we are currently exploring the possibilities of conducting such an analysis.

\begin{acknowledgements}

This project has received funding from the European Research Council (ERC) under the European Union's Horizon 2020 research and innovation programme (grant agreement 101020943, SPECMAP-CGM)  and from JSPS KAKENHI grant Nos. 23KJ2148 and 25K17444. This study made use of Python packages \texttt{astropy} \citep{astropy:2013, astropy:2018, astropy:2022}, \texttt{matplotlib} \citep{Hunter_2007_matplotlib}, \texttt{scipy} \citep{Scipy_2020}, \texttt{mpdaf} \citep{mpdaf_2017}, \texttt{numpy} \citep{numpy_2020}, \texttt{pandas} \citep{pandas_2010}, \texttt{seaborn} \citep{seaborn_2021}; as well as \texttt{galfit} tool \citep{Peng_2002}, \texttt{TOPCAT} tool \citep{topcat_2011}, \texttt{QFitsView} \citep{QFitsView_2012}, and \texttt{DS9} \citep{DS9_2000}. The sketch (Fig.~\ref{fig:summary_sketch}) was created using \texttt{https://www.canva.com} and \texttt{https://www.figurelabs.ai}. \texttt{Grammarly} tool (\texttt{https://www.grammarly.com}) was used for grammar assistance. 

\end{acknowledgements}

\bibliographystyle{aa} 
\bibliography{bibl.bib}

\begin{appendix}
\section{Selected LAEs: special cases}\label{ap:complex_laes}
For each of discussed below LAEs we provide $5\times 5 \arcsec$ pseudo NB Ly$\alpha$ images, extracted from MXDF in a way described in Sect.~\ref{section_data} and corresponding HST F775W images, provided in AMUSED database (\citealt{Bacon_2023}) (Fig.~\ref{fig:8537},~\ref{fig:7586}). Redshifts, overplotted in each panel, are spectroscopical $z_{\mathrm{spec}}$ for MXDF data and photometrical $z_{\mathrm{phot}}$ for HST data. Running AMUSED identification numbers are identified as 'ID' and running HST Rafelski catalog numbers as 'id' (\citealt{Rafelski_2015}).

\subsection{ID 8537}

     \begin{figure*}[h]
        \includegraphics[width=\hsize]{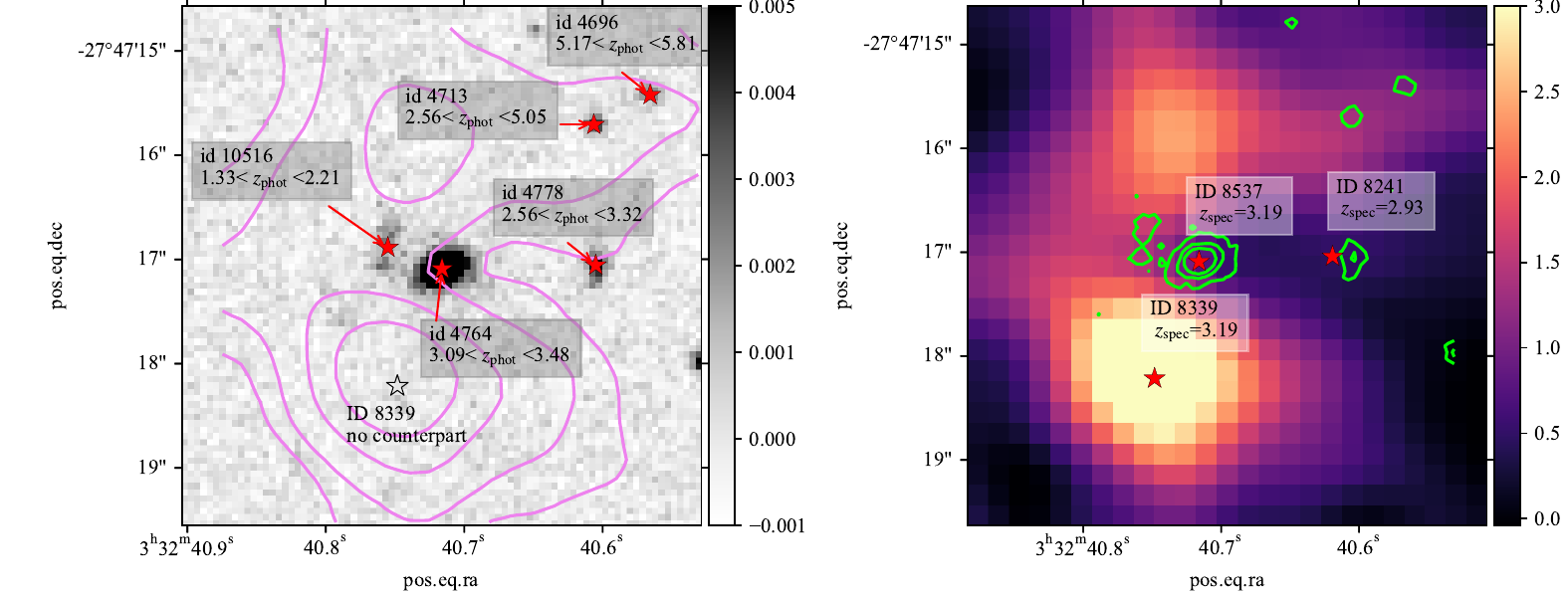}
        \caption{ID~8537: HST F775W image with (left) and NB Ly$\alpha$ image. We overplot contours in original resolutions from one image onto another to make visual comparison easier.}
        \label{fig:8537}
    \end{figure*}

We identify Ly$\alpha$ centroid of the bright halo around ID~8537 ($z_{\mathrm{spec}} = 3.18$)  to be the same as HST counterpart id~4764 ($3.09<z_{\mathrm{phot}}<3.48$) (Fig.~\ref{fig:8537}). It supported by the ID~8537 spectrum with its strong absorption lines \ion{Si}{ii}$\lambda 1260$, \ion{C}{ii}$\lambda 1334$, and \ion{Al}{ii}$\lambda 1671$ at the same $z_{\mathrm{spec}} = 3.18$, centered spatially on id 4764. ID~8339 at $z_{\mathrm{spec}} = 3.19$ is most likely not a galaxy itself, but rather a bright knot in the clumpy-like LAH of ID~8537. It has no HST counterpart in all available HST filters, provided in AMUSED (namely F435W, F606W, F775W, F850LP). Therefore, we assume Ly$\alpha$ emission being fainter in the center of the halo and brighter at larger distances, possibly due to radiative transfer effects. We model ID~8537 with \texttt{galfit} by masking out central $1.5\arcsec$ and using only one extended exponential component (see Sect.~\ref{section_data}). 

\subsection{ID 7586}

     \begin{figure*}[h]
        \includegraphics[width=\hsize]{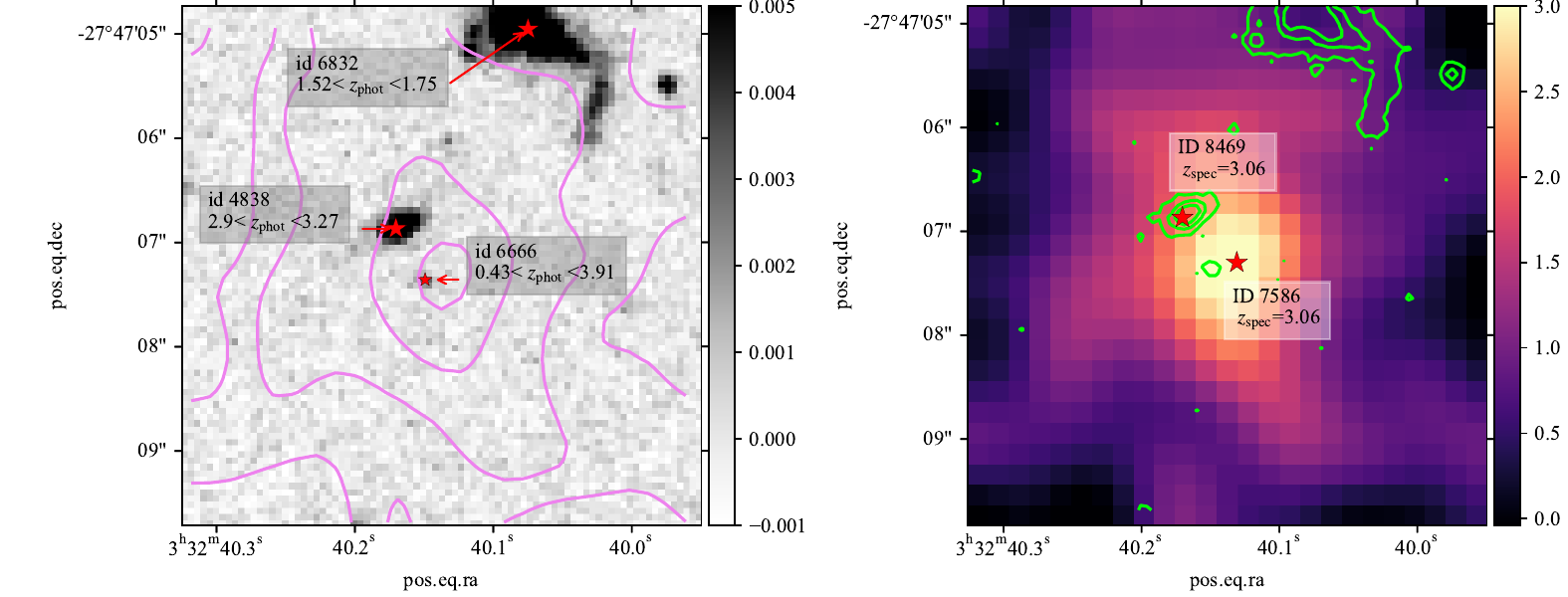}
        \caption{ID~7586: HST F775W image (left) and NB Ly$\alpha$ image. We overplot contours in original resolutions from one image onto another to make visual comparison easier.}
        \label{fig:7586}
    \end{figure*}

System ID~7586+8469 (Fig.~\ref{fig:7586}) consists of two LAEs ($r\simeq0.6\arcsec$) at the same redshift $z_{\mathrm{spec}} = 3.06$ both in AMUSED and in HST (id~6666+4838). Most probably both objects share the same LAH. Ly$\alpha$ emission was detected in ID~8469 in AMUSED and its centroid coaligns with the bright extended HST object (id~4838). However there is no strong spatial Ly$\alpha$ peak. At the same time ID~7586 has a point-like slightly offset HST counterpart (id~6666) but Ly$\alpha$ emission of this system corresponds to it.  There are two possible ways to interpret this system. First, that there is a LAH around ID~8469 but it does not contributes substantially to the whole system. Second, that detected Ly$\alpha$ emission in ID~8469 belongs rather to the LAH of ID~7586. In both cases it makes sense to produce \texttt{galfit} model only for ID~7586 as it gives the best representation of this system's spatial SB distribution and average SB radial profile (see Fig.~\ref{Fig_SB_profiles}).

\section{Recovering Ly$\alpha$ halo scalelengths}\label{ap:scalelengths}

     \begin{figure}[h]
        \includegraphics[width=\hsize]{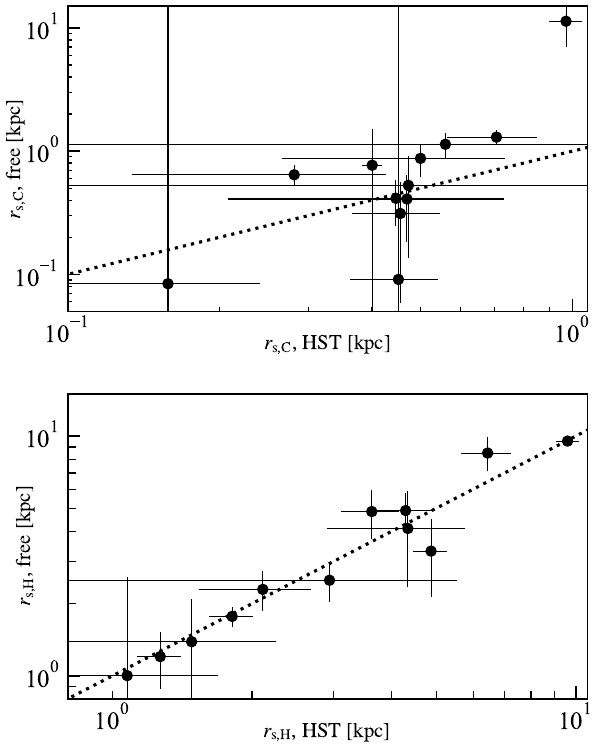}
        \caption{Top panel: exponential scalelength of central component $r_{s,C}$ obtained from HST versus $r_{s,C}$ obtained from the two-component model with scalelength as a free parameter. Bottom panel: exponential scalelength of an extended halo component $r_{s,C}$ obtained from the two-component model with fixed central parameters versus $r_{s,C}$ obtained from two-component model with free $r_{s, C}$. The dotted line indicated one-to-one correspondence}
        \label{Fig_rs_test}
    \end{figure}

 In this test we explore how fixed/unfixed scalelength in central component of two-component Ly$\alpha$ halo model influences a scalelength of the extended component. To achieve this, we produced two sets of models. First: central component with position angle $PA$, minor to major axis ratio $q$ and exponential scalelength $r_{s,C}$ as fixed HST parameters. Second: central component with $PA$ and  $q$ as fixed HST parameters and $r_{s,C}$ as a free model parameter. In both cases second extended halo component is enforced to be circular, but with $r_{s,H}$ as a free parameter. The results are shown in Fig.~\ref{Fig_rs_test}. Top panel shows a comparison between $r_{s,C}$ HST, obtained from HST counterpart, and $r_{s,C}$ free, as a result of the modeling with free scalelength. Dotted line is a unity line, and we see, that there is a scatter between fixed and free $r_{s,C}$. \texttt{galfit} tends to overestimate free $r_{s,C}$ in comparison to $r_{s,C}$ from HST. However, bottom panel shows, that even this mismatch does not lead to crucial differences in extended halo scalelengths, which is essential for our study. Both models provide very similar $r_{s,H}$. We conclude, that the scatter in $r_{s,C}$ is negligible to describe an extended component, as values of $r_{s,C}$ are typically $\le 0.5$ kpc, while $r_{s,H} > 1.5$ kpc.

\section{The contribution of undetected LAEs to the Ly$\alpha$ SB profile} 
\label{ap:SB_unb_calculation}

Here we summarize the formalism to predict the apparent surface brightness due to the integrated contribution of low-luminosity LAEs below the individual detection limit. We largely follow the procedure of \citet[][hereafter HA23]{Herrero_Alonso_2023}, except that here we perform this calculation for each object in our sample at its own redshift, whereas HA23 made a generic prediction at a nominal redshift of $z=3.5$. The expected integrated surface brightness $\text{SB}_{\text{unb}}$ from undetected neighboring LAEs is obtained as

    \begin{gather}
        \text{SB}_{\text{unb}} = 
        \int\limits_{z-\Delta z/2}^{z+\Delta z/2} \frac{V_c(z)\text{d}z}{4\pi d_L^2}
        \int\limits_{L_{\text{min}}}^{L_{\text{max}}}
        L\, \phi(L)\: \zeta(L, \Delta z) \: [1 - S(L, z)] \,\text{d}L
        \label{SB_laes}
    \end{gather}
    
where $S(L, z)$ is the LAE selection function in the MXDF, $\zeta(L, \Delta z)$ is the boosting factor due to clustering, $V_c(z)$ is redshift-defined volume, and $\phi(L)$ is the luminosity function of LAEs.

The mean number of faint LAEs is given by the Ly$\alpha$LF (more accurately, by an extrapolation of this distribution towards faint luminosities, see below). We adopt the usual prescription as a Schechter function \citep{Schechter_1976}: 
    \begin{gather}
        \phi(L)\text{d}L = \phi \left( \frac{L}{L^*}\right)^{\alpha}\text{exp}\left( - \frac{L}{L^*}\right)^{\alpha}\frac{\text{d}L}{L^*},
    \end{gather}
and assume the parameters from \citet{Tornotti_2025} based on the several MUSE surveys, including the deepest MXDF and MUDF, with a faint-end slope of $\alpha = -1.68$, a  characteristic luminosity of $\log L^* = 42.68$ $[$erg s$^{-1}]$, and a space density at $L^\star$ of $\text{log}\phi^* = -2.74$ $[\mathrm{Mpc}^{-3}]$. All parameters are assumed to be non-evolving between $z=2.9$ and 4.

The redshift-dependent flux or luminosity limit below which sources are no longer detectable is provided by the average MUSE/MXDF selection function $S$ for LAEs, which we construct using the method by \citet{Pharo_2024}. It is a bivariate function of flux and wavelength (or luminosity and redshift) that gives the probability ($0 \le S \le 1$) that an object of given properties is detected as a source by the survey. Here we are actually interested in the value of $1 - S(z)$ to get the number of \emph{undetected} LAEs. 

The clustering of galaxies enhances the number of neighbors to a given system, increasing their contribution to the apparent SB at large radii. To quantify this enhancement we use the halo occupation (HOD) modeling results by HA23 who expressed this effect in terms of a ``boosting factor'' $\zeta(L, r, \Delta z)$ depending on Ly$\alpha$ luminosity $L$, the distance $r$ to the central object, and the spectral width of the adopted narrowband filter (or pseudo-NB in the case of IFUs) converted into a redshift range. We estimate $\zeta$ for a fixed bandwidth of $\Delta z = 600$~km/s, small separations of $r < 0.1$~Mpc, and a dependency on $L$ based on a linear extrapolation of the trend found by  \citep{Herrero_Alonso_2023A}. This results in $\zeta \sim 16$ for marginally detected LAEs and $\zeta \sim 3$ at the extreme low-luminosity end of $\log L_{\text{min}} = 37$ $[$erg s$^{-1}]$ (see below), or approximately $\zeta \sim 4$ for the whole considered luminosity range.

The main uncertainties for Eq.~\ref{SB_laes} are the faint-end slope of the Ly$\alpha$ LF and the lower LF integration limit $L_{\text{min}}$. Note that the upper limit can be set to any large enough value as it is implicitly reset by the complement of the selection function $1-f_\mathrm{c}$ becoming zero. $L_{\text{min}}$, on the other hand, specifies the faintest LAEs that still contribute to the integrated luminosity density, with a weight given by the Ly$\alpha$LF. Some previous authors (\citealt{Bacon_2021}, HA23) used $\log L_{\text{min}} = 37$ $[$erg s$^{-1}]$, which is an extremely low limit that approximately corresponds to the emission from the \ion{H}{ii} region around a single O-type star. It turns out that given our adopted faint-end slope of the Ly$\alpha$LF the exact value of this lower limit does not matter much as the fractional contribution from such faint hypothetical LAEs is negligible. To estimate this breakdown by luminosity we calculate the differential number d$N$ of LAEs in a given differential co-moving volume as $\text{d}N = \phi(L)\,\text{d}L\,\text{d}V_c$, together with their corresponding contribution to the integrated surface brightness $\text{dSB} = \int_{z-\Delta z/2}^{z+\Delta z/2} \frac{\text{d}z}{4\pi d_L^2} L\times \text{d}N$. The resulting numbers increase with the size of the annulus adopted for the SB profile and also depend somewhat on redshift. An example outcome is given in Fig.~\ref{Fig_expected_N} where we step through the full luminosity range in increments of $\Delta \log L = 0.5$. 

\begin{figure}
\centering
    \includegraphics[width=\hsize]{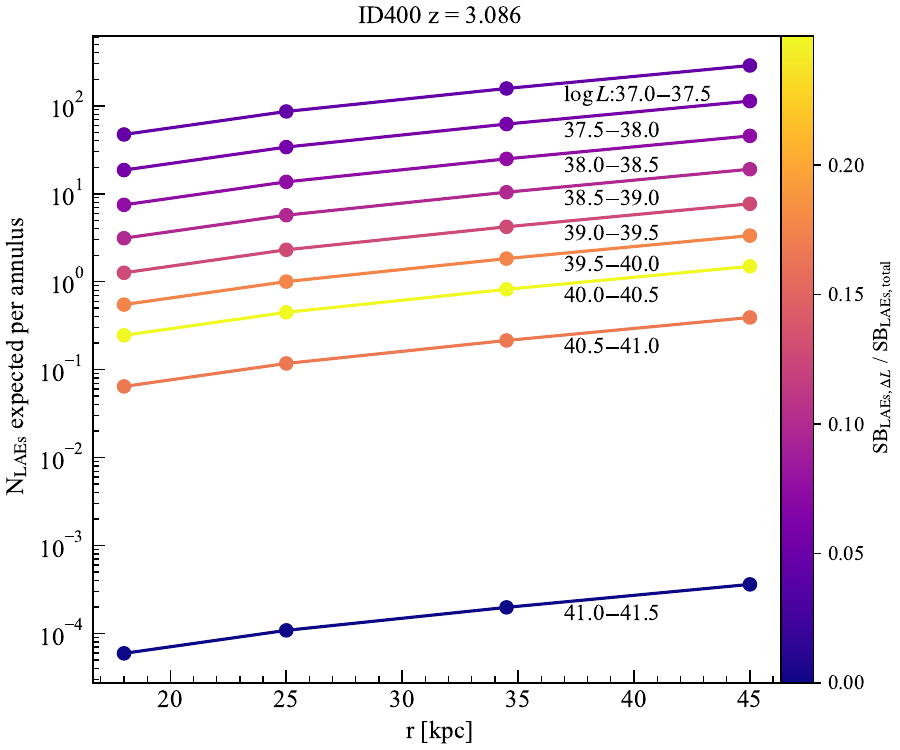}  
    \caption{Expected number of LAEs N$_{\text{LAEs}}$ per annular area, adopting the radial bins specified in Sect.~\ref{radial_profiles} and split into several LAE luminosity intervals $\Delta L$.  The colors correspond to the fractional contribution of these LAEs to the total integrated SB$_{\text{unb}}$ of undetected LAEs. The estimated values include the MXDF selection function and a luminosity-dependent enhancement factor for clustering (taken from HA23). We show these numbers for object ID~400 as an example, but the results are similar for all objects in our sample. 
    }
\label{Fig_expected_N}
\end{figure}

According to Fig.~\ref{Fig_expected_N}, $>70$\% of the integrated surface brightness is contributed by LAEs with $39 < \log{L} < 40.5$ $[$erg s$^{-1}]$. Based on these estimates, we keep the lower integration limit of $\log L_{\text{min}} = 37$ $[$erg s$^{-1}]$ specified above. The figure also shows that LAEs in the luminosity range that contributes most are actually still quite rare within such a small annular area, and any individual occurrence would produce a lopsided appearance of the halo. Note that the numbers in Fig.~\ref{Fig_expected_N} also include the above defined clustering enhancement factor $\zeta(L,z)$, taken from HA23.

\section{Ly$\alpha$ surface brightness from UVB fluorescence}
\label{ap:uvb_calculation}

We follow \citet{Gallego_2021} in most aspects, except that we perform the calculation separately for each object at its actual redshift and use the updated UVB model of \citet{Puchwein_2019}. To summarize the main assumptions \citep{Cantalupo_2005}, each CGM cloud is taken as a spherical cloud of hydrogen in ionization equilibrium with a column density $\text{N}_{\text{\ion{H}{i}}} > 10^{17.2} \text{ cm}^{-2}$. We assume case B recombination and a temperature of $T = 2 \times 10^4$~K, the expected temperature for photionized gas in thermal equilibrium \citep{Cantalupo_2005}, implying that about 65\% of all recombinations lead to the emission of a Ly$\alpha$ photon. The Ly$\alpha$ surface brightness of a system of clouds is then given by 

    \begin{align}\label{Eq_SB_uvb}
        \text{SB}_{\mathrm{Ly}\alpha} \simeq 
        & 1.84\times \text{erg}\:\text{cm}^{-2}\:\text{s}^{-1}\:\text{arcsec}^{-2} \:
        \nonumber\\
        & \times\left(\frac{\Gamma_{\ion{H}{i}}}{10^{-12}\,\text{s}^{-1}}\right) \times
        \left(\frac{\bar{\sigma}_{\ion{H}{i}}}{10^{-18}\,\text{cm}^{2}}\right) \times
        (1+z)^4 \times f_{\text{c}} \:.
    \end{align}

The frequency-averaged photoionization cross-section $\bar{\sigma}_{\ion{H}{i}}(z)$ depends only on the spectral shape of the UVB, which in the fiducial UVB model of \citet{Puchwein_2019} does not vary significantly with $z$, and we therefore take its value as a constant, $\bar{\sigma}_{\ion{H}{i}}(z)\simeq 2.77\times 10^{-18}$~cm$^{2}$. We obtain the photoionization rate $\Gamma_{\ion{H}{i}}(z)$ from \citet{Puchwein_2019} by interpolation between the tabulated values for different redshifts.

We are also interested in the local photoionization rate due to massive stars in the central galaxy. At distance $r$ to the galaxy this is 

    \begin{gather}
        \Gamma_{\ion{H}{i}, \text{ local}}(r) \approx f_{\text{esc, LyC}}\,\frac{R \, \bar{\sigma}_{\nu_{\ion{H}{i}}}}{4\pi r^2} \:
        \label{Eq_gamma}
    \end{gather}

where $R$ is Ly$\alpha$ photon production per unit time, surface and solid angle. The distance from the galaxy where $\Gamma_{\ion{H}{i}, \text{ local}}(r)$ becomes equal to that due to the metagalactic UVB is then

    \begin{gather}
        r_{\text{eq}} = \sqrt{f_{\text{esc, LyC}}\,\frac{R \,\bar{\sigma}}{4 \pi \,\Gamma_{\ion{H}{i}, \text{ UVB}}}} \:.
    \end{gather}

\end{appendix}

\end{document}